 \documentclass[12pt]{article} 
\usepackage{amsmath,amssymb}   
  
\catcode`@=11
\def\secteqno{\@addtoreset{equation}{section}%
\def\theequation{\thesection.\arabic{equation}}}
\catcode`@=12
\secteqno 
\newcommand{\be}{\begin{equation}}
\newcommand{\ee}{\end{equation}}
\newcommand{\bea}{\begin{eqnarray}}
\newcommand{\eea}{\end{eqnarray}}
\newcommand{\bref}[1]{(\ref{#1})}
\newcommand{\nn}{\nonumber}
\newcommand{\A}{\alpha} \newcommand{\B}{\beta} \newcommand{\gam}{\gamma}
 \newcommand{\D}{\delta} 
\newcommand{\ep}{\epsilon} 
 
   \newcommand{\vp}{\varphi}
\newcommand{\lam}{\lambda}      \newcommand{\s}{\sigma}
          \newcommand{\w}{\omega}
          
\newcommand{\h}{\eta}           
           
\newcommand{\W}{\Omega}         
            
 \newcommand{\Lam}{\Lambda}

\newcommand{\ba}{\overline }
\def\6{\partial} \def\7{\tilde} \def\8{\hat}

\def\pa{\partial}

\def\CL{{\cal L}}
\def\CH{{\cal H}}\def\CK{{\cal K}}

\def\l{{\ell}}\def\CR{{\cal R}}
\def\vs{\vskip 3mm}\def\={{\;=\;}}\def\+{{\;+\;}}

\def\too{\quad\to\quad}
\def\ua{{\underline a}} \def\ub{{\underline b}} \def\ud{{\underline d}} 
\def\uc{{\underline c}} \def\vp{\varphi} 
\def\lag{Lagrangian}
\def\ebox#1#2{\vskip 2mm{\vbox{\hrule\hbox{\vrule\kern3pt\vbox{\kern3pt
         {\begin{eqnarray}#1\label{#2}\end{eqnarray}}
         \kern3pt}\kern3pt\vrule}\hrule}}\vskip 2mm}
\def\tbox{\vskip 2mm{\vbox{\hrule\hbox{\vrule\kern3pt\vbox{\kern3pt
         {{\hfill {\small ${}^{notebook\; kiyoshi
         }$} \\
         \large \bf ~~\reptitle}\\ } 
         \kern3pt}\kern3pt\vrule}\hrule}}\vskip 2mm}
\def\lag{Lagrangian }
\def\vs{\vskip 4mm}

\begin{document}

{\hfill {\rm  UB-ECM-PF-12-74 , ICCUB-12-157}}
\vskip 2cm 
\begin{center} 
{\Large \bf Non-linear Realizations, Goldstone bosons of broken Lorentz rotations and effective actions 
for p-branes}
\vskip 15mm 
{\large
 {Joaquim Gomis${}^a$} and {Kiyoshi Kamimura${}^b$}, {and Josep M. Pons${}^a$}
}\vskip 1cm
{\it 
{${}^a$  { Departament d'Estructura i Constituents de la Mat\`eria and Institut de Ci\`encies del Cosmos, 
Universitat de Barcelona, Diagonal 647, 08028 Barcelona, Spain}
}

{${}^b$ Department of Physics, Toho University, Funabashi, Chiba 274-8510, Japan}
\vskip 5mm

{\it E-mails:  gomis@ecm.ub.es, kamimura@ph.sci.toho-u.ac.jp, pons@ecm.ub.es, } 
}
\end{center}
\vskip 3cm

\abstract
{
We consider the non-linear realizations of the Poincare group for $p$-branes with  
local subgroup $SO(1,p)\times SO(D-(p+1))$.  The Nambu-Goto $p$-brane action is 
constructed using the Maurer Cartan forms of the unbroken translations. 
We perform a throughout phase space analysis of the action and show that it leads 
to the canonical action of a $p$-brane.  We also construct   some higher order 
derivative terms  of the effective  $p$-brane action using the MC forms of the broken 
Lorentz transformations.}

\vskip 10mm
keywords:  {Non-linear Realizations, p-branes, Space-Time Symmetries,
Effective Actions}

\eject

\section{Introduction \label{sec:0}}

Recently there has been a renewal of interest to construct the effective string theory action 
\cite{Luscher:2004ib} (see also \cite{Polchinski:1991ax})
using non-linear Lorentz invariance. The lowest order term in derivatives  is the well  known 
Nambu-Goto (NG) action. 
In  \cite{arXiv:1111.5758},  and references therein, it has been examined the 
next order Lorentz invariant corrections to the NG action in the static 
gauge\footnote{ See also  \cite{Billo:2012da},\cite{Dubovsky:2012sh} for recent 
work on effective strings in the static gauge.}.  
In that approach it is required the invariance under the broken Lorentz transformations of 
coordinates of the string, which are the Goldstone bosons associated with the broken translations. The 
broken Lorentz transformations rotate  
the longitudinal and transverse directions of the string.
In this paper we consider $p$-brane diffeomorphism (Diff) invariant actions in the non-linear 
realization (NLR) approach \cite{Coleman}, see also  \cite{ogi}.  In our case they contain the Goldstone bosons of both 
broken and unbroken translations and Lorentz rotations. The MC forms, pullbacked to the world volume, 
are used to build these Lagrangians with the geometric tools of the wedge product, Hodge operator, and covariant 
differentiation. These actions are  functionals
 of both of the embedding variables $x$ 
and the Lorentz variables $\Phi$.


The general action will be
\be
S[x,\Phi]=\int e[x,\Phi] {\cal L}[x,\Phi]\,,
\label{both}
\ee
where the volume form $e[x,\Phi]$ is constructed by using the Maurer 
Cartan (MC) forms associated to unbroken translations. With ${\cal L}=1$, it leads to the NG 
action for the embedding variables, see for example 
\cite{Ivanov:1999fwa}\cite{Bellucci:2002ji}\cite{Gomis:2006xw}\cite{hep-th/0607104},
 once the Inverse Higgs (IH) mechanism \cite{Ivanov:1975zq}, or 
equivalently the equations of motion (EOM) of the Lorentz variables are used\footnote{For a recent discussion about the relation between the IH mechanism and the equation of motion of non-dynamical Goldstone bosons see \cite{McArthur:2010zm}.}. In this lowest order case 
in derivatives, the 
Lorentz variables are non-dynamical.

To construct more general actions with ${\cal L}\neq 1$  we can use the 
MC forms associated to the broken  and the unbroken Lorentz transformations. In this case the
Lorentz variables are dynamical. Note that the form of the invariant 
scalar Lagrangian has no additional restrictions, and there is an infinite arbitrariness of choices.
 For $p$-brane actions we should also consider WZ terms $ {\cal L}_{WZ}$
that are constructed
from closed invariant p+2 forms, $\Omega_{p+2}=d{\cal L}_{WZ}$.

If we want to obtain a Lagrangian depending on the embedding coordinates $x$'s alone, we should express the Lorentz variables in terms of  these coordinates. 
A standard procedure is to impose an invariant set of constraints, so called 
the IH mechanism (see \bref{IH00}) 
and to construct 
the corresponding MC forms on the world volume associated to the broken Lorentz transformations and to the  unbroken rotation (to build 
covariant derivatives) and use them to construct the polynomial and non-polynomial invariants 
of the coefficients of these forms along the lines of, for example \cite{ogi}.
This procedure give all possible local terms in the effective action.

In this paper we explore a possible different method to express 
the Lorentz variables  in terms of geometrical quantities by solving the equations of motion perturbatively.    
A starting Lagrangian is called  
the {\sl seed} Lagrangian in the following sense:
since  we want to obtain a Lagrangian depending on the embedding coordinates $x$'s alone, we should express the Lorentz variables in terms of these variables. This can be done by solving perturbatively the equations of motion of the Lorentz variables and substitute these perturbative solutions into the {\sl seed} Lagrangian. Thus from a {\sl seed} Lagrangian we obtain a perturbative series of Lagrangians in terms of the embedding variables. 
Let us insist however that there is an infinite freedom of choosing the seed Lagrangian. Obviously we could limit ourselves to  polynomial seed Lagrangians, organized in number
of derivatives,  as is done in effective theories.

\vs
 
We first illustrate   it here in the case of a particle (see details in the section 3). 
We consider a coset $G/H=({\rm {Poincare}})/SO(D-1)$ and the coset element   
$g= g_0\,U, \label{gold20}$ where $g_0=e^{i x^0 P_0}e^{i x^{a'}P_{a'}} $
is the coset representing the Minkowski space and $U=e^{i {\theta}_{0}{}^{b'} J^{0}{}_{b'}}$ is a general Lorentz rotation, $x^{a'}\, (a'=1,...D-1)$ are the Goldstone bosons
of broken translations, $x^0$ is the Goldstone boson of the unbroken time translation\footnote{The unbroken Lorentz translation $P_0$ generates via a right {action}  \cite{McArthur:1999dy}\cite{hep-th/0607104} a transformation which is equivalent to the world-line diffeomorphism.} and  $U$ is parametrized by the Goldstone bosons of the broken Lorentz transformations.
The MC forms are given by, $(\ua=0,...,D-1)$\footnote{We follow the notations of  \cite{Gomis:2006xw}.}, 
\bea
\W&=&-ig^{-1}dg=L^{\ua}P_{\ua}+\frac 12 L_{\ua}{}^{\ub}\,J^{\ua}{}_{\ub},
\qquad L^{\ub}=dx^{\ua}{\Phi_\ua}^{\ub},\quad L^{\ua\ub}=(\Phi^{-1} d\Phi)^{\ua\ub},
\label{MC-forms}
\eea
where ${\Phi_\ua}{}^{\ub}(\theta)$ is a Lorentz boost parametrized by 
${\theta}_{0}{}^{b'}$. The action\footnote{In writing the Lagrangians, the pullback to the world volume of the MC forms is always understood. The pullback notation ${}^*$ is sometimes suppressed just for notational simplicity.} of a free massive particle
is the  $SO(D-1)$ invariant one form $L^0$  associated to the unbroken translation $P_0$,
\be
I=-m\int(L^0)^*
=-m\int d\tau \,\dot x^\ua {\Phi_\ua}^0(\theta)=-m\int d\tau \,{e}.
\ee
This action is invariant under Diff, as is the case for every action built from the pullback of MC forms and with the tools mentioned above. If we compute the momenta  
\be
p_\ua=\frac{\pa \CL}{\pa\dot x^\ua}= -m{\Phi_\ua}^0
\label{pPhipa}\ee
therefore ${\Phi_\ua}^0(\theta)$  
are functions of the momenta of the particle. { This fact implies that the Goldstone bosons of broken rotations
${\theta}_{0}{}^{b'}$ are actually phase space degrees of freedom of the particle}.

Now if we regard ${\Phi_\ua}^0$ as independent degrees of freedom 
we can rewrite the action as 
\be 
I=-m\,\int d\tau\left(\dot x^\ua {\Phi_\ua}^0+\frac{\gamma}2(\h^{\ua\ub}{\Phi_\ua}^0{\Phi_\ub}^0+1)\right)=
\int d\tau\left(p_\ua\dot x^\ua -
\frac{\gamma}{2m}(\h^{\ua\ub}{p_\ua}{p_\ub}+m^2)\right),
\label{particlepaction}
\ee
where $\gam$ is the Lagrange multiplier to constrain ${\Phi_\ua}^0$ to be a time-like unit Lorentz vector. 
It is the canonical action of the relativistic free particle, as can be extracted from \cite{Casalbuoni:2008iy}  \cite{Gibbons:2009me}. Therefore the NLR action that include the Goldstone bosons of the translations and  the Goldstone bosons associated to the broken 
boost is the canonical action of the massive relativistic particle.

If we use the equations of motion of ${\Phi_\ua}^0$ and $\gamma$, we obtain
\be\label{inversehiggs}
{\Phi_\ua}^0=-\frac{\dot x_\ua}{\sqrt{-\dot x^2}}.
\ee
Note that ${\Phi_\ua}^{a'}$ is normal to the worldline and
 is equivalent to the vanishing the MC form $L^{a'}$ associated to the broken translations
\be
L^{a'}=dx^{\ua}{\Phi_\ua}^{a'}=0
\label{IH00}
\ee
which is known as the inverse Higgs (IH) mechanism \cite{Ivanov:1975zq}.
\vs

We can construct corrections to the NG action of the particle by considering  
the pullback of the MC forms associated to the
broken Lorentz transformations\footnote{ In \cite{Gauntlett:1989qe}  \cite{Gauntlett:1990nk} the MC forms associated to broken Lorentz rotations were used to construct relativistic  {gauge fixed} (super) particle action in  two dimensions with extrinsic curvature. }.
We use the standard tools (wedge product, Hodge operator, and covariant differentiation) available with the MC forms.  In the case of the particle we define the Hodge operator
by assigning  $L^0=d\tau\,e$  as the volume form which implies $ *d\tau=-{e^{-1}}$. Since we want corrections to the NG action in the spirit of effective theories, we first consider the 
term with first order derivatives of the Lorentz variables. In particular the polynomial one, which is quadratic in the velocities, whereas the coefficient is chosen to be small. 
This is made by choosing, among the $SO(D-1)$  invariant one forms, the structure $L_{0a'}*L^{0a'}$, which depends at most on the velocities $\dot\Phi^{\ua\ub}$. 
Adding this term to the action \bref{particlepaction} we obtain
\be\label{effective00}
I=-m\int d\tau\{\,e+\frac{\gamma}2({\Phi_\ua}^0{\Phi^{\ua 0}}+1)+
\frac \B{2m^2}\,\left(
\frac{1}{e}\,{{\dot\Phi}^\ua}{}_0(\h_{\ua\ub}-\Phi_{\ua 0}{\Phi_{\ub}}^{0}){\dot\Phi}^{\ub 0}\right)\},
\ee
or better, with a redefinition of $\gamma$,
\be\label{effective0}
I=-m\int d\tau\{\,e+\frac{\gamma}2({\Phi_\ua}^0{\Phi^{\ua 0}}+1)+
\frac \B{2m^2}\,(
\frac{1}{e}\,{{\dot\Phi}^\ua}{}_0{\dot\Phi}^{\ua 0})\},
\ee 
where the constant parameters $\beta$ is dimensionless and the term with $\beta$ is considered as a perturbation of the free particle action\footnote{We could have also added terms with higher derivatives in $\Phi$. We will not consider those terms in this paper.}.
Notice that in this Lagrangian the $ {\Phi_{\ub}}^{0}$  variables are independent dynamical variables. 
We have been unable to solve dynamically  to all orders in $\beta$ the equation of motion of 
$ {\Phi_{\ub}}^{0}$ in terms of $x^\ua, \gamma$.

We do not use other possible Lagrangian candidates, like $L_{a'}*L^{0a'}$ or $L_{a'}*L^{a'}$, since they vanish at lowest order due to \bref{IH00}. We could also add invariant terms with an increasing number of derivatives of $ \Phi $ like for instance $ (\tilde D*L_{0a'})*(\tilde D*L_{0a'})$, where $\tilde D$ is the covariant differential,  as in effective theories. We will focus in this paper the lowest order in derivatives,  and we will use this action (\ref{effective0}) as a seed to produce a set of geometrical invariant actions. In fact
in order to make contact with an effective Lagrangian written in terms of the geometric quantities of the
world-line in Minkowski space we use the procedure of solving the equations of motion of
${\Phi_{\ub}}^{0}$ perturbatively in $\B$, in terms of the velocities, accelerations,... of the particle.
At lowest order in $\beta$, $ {\Phi_{\ub}}^{0}$ is given by \bref{inversehiggs} 
and the geometrical action becomes\footnote{If we only consider the  term of order $\beta$ the dynamics of that Lagrangian is not equivalent to the starting \lag \bref{effective0}. The equivalence would only be achieved with the full perturbative series.}, in terms of proper time $s$, 
 \be
I=-m\int ds \Big(1+\frac{\B }{2m^2}\kappa_1^2+O(\B^2)\Big)
\ee
where $\kappa_1$ is the first curvature of the world-line, 
$\kappa_1^2=(A)^2,\; A^\ua=\frac{d^2x^\ua}{ds^2}$. Note that we do not make use of the EOM of the coordinates $x$, unlike what is customarily done in the standard procedure for effective field theories \cite{Polchinski:1991ax}\cite{arXiv:1111.5758}.
At first order in $\B$ the solution for ${\Phi_{\ub}^0}$ from (\ref{effective0}) departs from the IH mechanism,
\be
{\Phi_\ua}^0=-U_\ua-\B \Sigma_\ua
\label{phibeta}
\ee
where $U_\ua=\frac{d x_\ua}{ds}$ and $\Sigma_\ua=\frac{d^3x_\ub}{ds^3}({\D_\ub}^{\ua}+U_\ub U^{\ua})$ is the {\it relativistic jerk vector}, see for example \cite{Dunajski:2008tg} \cite{Russo:2009yd}. 
Note that there is a rationale for the presence of the relativistic jerk vector, since it is orthogonal to the velocity, and this orthogonality is already imposed on ${\Phi_\ua}^0$ in (\ref{phibeta}) due to the Lorentz constraints $\h^{\ua\ub}{\Phi_\ua}^0{\Phi_\ub}^0+1=0$.

By substituting \bref{phibeta} into \bref{effective0} we can
consider the expansion of the action up to order $\beta^2$ in terms of the embedding variables,
\be
I_2=m\,\int ds\,\frac{\beta^2}{2m^4}\,(\kappa_1^2 \kappa_2^2+
(\frac{d\kappa_1}{ds})^2),
 \label{paeffb2}\ee
 where $
  \kappa_2^2=(\frac{d^3x}{ds^3} O_2 \frac{d^3x}{ds^3})/\kappa_1^2 $
 is the second curvature of the world-line,
 $({O_2})$ is the projector orthogonal to accelerations and velocities ($s$ is proper time). 
  We should notice that the combination of curvatures  $ \kappa_1^2 \kappa_2^2+
(\frac{d\kappa_1}{ds})^2$ 
appearing at order $\beta^2$ of the effective \lag \bref{paeffb2}
 is the $\Sigma^2$ of the relativistic jerk vector $\Sigma_{\ua}$ . 

Note that we could consider within our approach more general, non-polynomial,  actions, like
\be \CL=-e\left(1+ \frac{\gamma}{2}(\Phi^2+1)\right)+\frac{\tilde\B}2{\sqrt{(\dot\Phi)^2}}.
\label{Lgam} 
\ee
In this case the relation to order $\tilde\B$ among the Lorentz variables and  the geometrical quantities is given 
\bea
\Phi^{\ua\,0}&=&
- U^\ua+\frac{\tilde \B}{2\sqrt{-\dot x^2}}\frac{ \,A^{\ua}}{ \sqrt{(\,A)^{2}}}
\eea
As long as the seed Lagrangian contains the term with first time derivatives of the Lorentz variables, the first correction term to the NG will be a functional of the first extrinsic curvature.

\vs


In this paper we will see that the previous results for the particle are extended to the case of a $p$-brane and show that: 

\begin{enumerate}

\item
The Goldstone bosons of broken Lorentz rotations are functions of the phase space 
degrees of freedom of the $p$-brane.

\item
The NLR action constructed from the MC forms of unbroken translations leads
in a natural way to the canonical  form of  NG action.  
Since the Goldstone bosons of the broken rotations are non-dynamical we can eliminate them through their own equations of motion and one gets the world-volume NG action  \cite{Gomis:2006xw}.

\item
We can also construct a WZ term for a $p$-brane
in dimensions $p+2$ using only the MC forms of the translations\footnote{For the case of the string see  \cite{Ramos:1997fa}\cite{Curtright:2010zz} and for the 2-brane \cite{Bonanos:2008kr}.}.

\item
As in \bref{particlepaction} it is often useful to introduce also Goldstone bosons associated to the unbroken Lorentz rotations as independent
degrees of freedom without using explicit parametrization of the Lorentz transformation. 
In this case the Poincare invariant $p$-brane action will be invariant under local  $SO(1,p)\times SO(D-(p+1))$ rotations  as well as world-volume diffeomorphism (Diff). 

\item
  We also consider terms of the effective action of a p-brane with
non-linearly realized Poincare symmetry
  by adding invariant forms associated with the broken Lorentz transformations.
In these cases actions, like the particle case,  will contain the first derivatives 
of the Goldstone bosons of broken rotations and the ${\Phi}$'s become dynamical. 
The Lagrangian with both the embedding variables and the Lorentz variables is used as a {\sl seed}
to produce corrections, in terms of the embedding variables, to the NG Lagrangian. The procedure consist in solving for the Lorentz variables by using the equations of motion (EOM) iteratively and substituting them back into the \lag so that it will produce an effective action in terms of geometrical quantities, like the intrinsic curvature, 
extrinsic curvature and the higher curvatures of the world-volume.
\end{enumerate}

The organization of the paper is as follows.  In section 2 we will construct the action of a tensionfull $p$-brane with the lowest order derivatives using the non-linear realization approach with local subgroup $SO(1,p)\times SO(D-(p+1))$.  
We will perform the Hamiltonian-Dirac analysis that   
leads to the canonical action of the  NG action. In section 3 we will
construct leading order -and beyond- corrections to the NG action. 
Finally we will present some conclusions and outlook. There are two appendices with technical details.
\vs

{ Note added: when this paper was in a process of submission 
it has appeared a paper by Gliozzi and Meineri \cite{Gliozzi:2012cx} 
which proposes a different  procedure 
to construct some higher order derivative corrections to the
Nambu Goto action. Their results agree with ours. 
}
\section{Nonlinear realization with local subgroup \\ $SO(1,p)\times SO(D-(p+1))$}

In this section we will construct the $p$-brane action using the NLR approach  \cite{hep-th/0607104} by Callan, Coleman, Wess and Zumino   \cite{Coleman}. 
However we will not follow the  method in terms of coset representatives. 
In  the present case it is more  useful to reformulate the theory of non-linear realizations in terms of the group elements\footnote{For the case of internal symmetries see \cite{DPNU-87-63}.} of $G$ themselves rather than those of the coset $G/H$. We consider group elements  $g$ of $G$ and take the symmetries of the non-linearly realized  theory given by
\be
g(x)\to g_0 g(x) \ \ {\rm and }\ \  g(x)\to g(x) h(x)
\ee
where $g_0\in G$ is a rigid transformation while  the second independent local transformation $h(x)$ is an arbitrary space-time dependent transformation that belongs to $H$.
In this more general way 
the theory contains gauge degrees of freedom that  can be fixed using the local $H$ transformations. We will refer to such a formulation as a non-linear realization of a group $G$ with local subgroup $H$.

 The global $G$ transformations of the coset coordinates $x$'s are 
determined up to the local $H$ transformations and are in general non-linear.
One could from the beginning  use these local $H$ transformations to 
 fix the gauge freedom and so work only with coset representatives. This is equivalent to the original approach and it requires the $H$-compensating transformations 
for the global $g_0$ transformations. 
One can also work in a half way house where only some of the fields are removed.

For the case of relativistic $p$-branes we consider the NLR of $G={ISO(D-1,1)}$
with the local  subgroup $SO(1,p)\times SO(D-(p+1))$.
 The generators\footnote{$ \underline a$ goes over all possible values from $0$ to $D-1$ while the 
unprimed indices $a$ take the values  $a=0,\ldots , p$ and primed indices $a'$ take the values $p+1,\ldots , D-1$. The latter are the indices which are longitudinal and transverse to the brane respectively. } 
$  P_a , J^{a}{}_{b}, J^{a'}{}_{b'}$ generate the unbroken translations, local longitudinal and transverse rotations via the right transformations \cite{McArthur:1999dy}.
In the non-linear realization  the reparametrization invariance 
corresponds to the invariance under unbroken translations  
with additional local Lorentz rotations for a general $p$-brane, see \cite{hep-th/0607104}.

The group elements are parametrized by
\be
g=e^{i x^a P_a}e^{i x^{a'}P_{a'}}e^{i {\theta}_{a}{}^{b'} J^{a}{}_{b'}}
e^{i {\frac12}{\theta}_{a}{}^{b} J^{a}{}_{b}}e^{i {\frac12}{\theta}_{a'}{}^{b'} J^{a'}{}_{b'}}=g_0\,U,
\label{gold2}
\ee
where $g_0=e^{i x^a P_a}e^{i x^{a'}P_{a'}} 
$ is the coset representing the Minkowski space and $U$ is a general element of the Lorentz group.
In \bref{gold2} we have introduced ${SO(1,p)\times SO(D-(p+1))}$ degrees of freedom  in addition to the coset elements of $\frac{ISO(D-1,1)}{SO(1,p)\times SO(D-(p+1))}$ used for example in \cite{Gomis:2006xw}.
The MC form is given by
\be
\W=-ig^{-1}dg=L^{\ua}P_{\ua}+\frac 12 L_{\ua}{}^{\ub}\,J^{\ua}{}_{\ub},
\ee
where the vielbein  and spin connection forms in the flat $D$ dimensional target space are 
\bea\label{vielbein}
{L}^{\ua}&=&dx^\ub \,{\Phi_\ub}^{\ua}, \qquad L^{\ua}{}_{\ub}={\Phi_\uc}^{\ua}d{\Phi^\uc}_{\ub}, 
\eea
and are satisfying the MC equations
\be
dL^\ua+{L^\ua}_\ub\wedge L^\ub=0,\qquad dL^{\ua\ub}+{L^\ua}_\uc\wedge  L^{\uc\ub}=0.
\ee
Here ${\Phi_\ub}^{\ua}$ is a finite Lorentz transformation  parametrized
by ${\theta}_{a}{}^{b'}, {\theta}_{a}{}^{b}, {\theta}_{a'}{}^{b'}$.
The explicit form of the finite Lorentz transformation $\Phi$ in terms of the Goldstone fields ${\theta}$ is not simple.
Also, the local $SO(1,p)\times SO(D-(p+1))$ transformations of  ${\theta}$ are involved. 

{Like in the case of the particle \bref{particlepaction}}, it is more convenient to consider the $D^2$  elements of  ${\Phi_\ua}^\ub$ of the Lorentz transformation themselves as new variables restricted by the orthonormality constraint
\be\label{lorentz}
C^{\uc\ud}=\h^{\ua\ub}\,{\Phi_\ua}^\uc\,{\Phi_\ub}^\ud
-\h^{\uc\ud}=0\,,
\ee
and with the implicit choice of $\det{\Phi}=1, {\Phi_0}^0\geq1$.
Under the local rotations ${O_b}^a\in SO(1,p)$ and ${O_{b'}}^{a'}\in 
SO(D-(p+1))$ it transforms  
as
\be
{\Phi_\ua}^a\to {\Phi_\ua}^b {O_b}^a(\xi),\qquad
{\Phi_\ua}^{a'}\to {\Phi_\ua}^{b'} {O_{b'}}^{a'}(\xi),
\label{gaugeact}
\ee
where $\xi^i, (i=0,1,...,p)$ are the $p+1$ coordinates of the world-volume of a $p$-brane.
The number of Goldstone bosons $\theta$  of \bref{gold2} is $\frac{D(D-1)}{2}$. We describe them by the   
$D^2$ elements of  ${\Phi_\ua}^\ub$ 
 subject to the $\frac{D(D+1)}{2}$ conditions  \bref{lorentz}.
 $\frac{(p+1)p}{2}$ of  them  are the  gauge degrees of freedom of $SO(p,1)$
and $\frac{(D-(p+1))((D-(p+2))}{2}$ are the ones associated to the local  $SO(D-(p+1))$. So the number of non-gauge degrees of freedom for the $\Phi$ variables will be $(p+1)(D-(p+1))$
which agrees with the counting of Goldstone bosons ${\theta}_{a}{}^{b'}$ associated to broken Lorentz rotations of the coset $\frac{ISO(D-1,1)}{SO(1,p)\times SO(D-(p+1))}$  \cite{Gomis:2006xw}.

\vs

\subsection{ NG $p$-brane Lagrangian }

The action  of  relativistic $p$-branes  {with lowest order of derivatives} is constructed
from the pullback of the MC forms \bref{vielbein} on the world-volume, with dimensionless coordinates  $(\xi^0,\xi^m),  m=1,...,p$.
It must be invariant under world-volume diffeomorphism
associated to the longitudinal unbroken translation invariance 
and local longitudinal and transverse Lorentz rotations.  
The \lag density,  for a generic dimension,  is given by
\be\label{invariantdensity}
\CL=\CL_0+\CL_1,
\ee
with
$\CL_0$ is the world-volume density of the $p+1$-brane\footnote{We often omit the $\wedge$ symbol for the wedge products.} 
\bea
d^{p+1}\xi\,\CL_0 &=&
\kappa\, \frac{1}{(p+1)!}\ep_{a_0...a_p}e^{a_0}...e^{a_p}
=-d^{p+1}\xi\, \kappa\,e, 
\label{L0action}\eea
where $\kappa$ is the tension of the $p$-brane with dimension $\left[\kappa\right]={{m}}^{p+1}$ and
$e=\det({e_i}^a) $ with $(p+1)
$-bein given by
\be\label{vielbein0}
e^a=(L^a)^*=d\xi^i{e_i}^a\,\qquad {e_i}^a=\pa_i x^\ua \,{\Phi_\ua}^a
\ee 
which is the pullback of the 
$L^a$ in \bref{vielbein}.
 $\CL_1$ is the Lagrange multiplier term for the orthonormality 
of the  $D$-beins ${\Phi_\ua}^\uc$, \bref{lorentz},
\be \CL_1=e\,\frac{\gam_{\uc\ud}}2(\h^{\ua\ub}\,{\Phi_\ua}^\uc\,{\Phi_\ub}^\ud
-\h^{\uc\ud}). 
\ee
We can write the Lagrangian \bref{invariantdensity} as\footnote{A similar action in terms of Lorentz harmonics was considered in \cite{Bandos:1992iu}  \cite{Bandos:1995zw}.}
\be\label{final}
\CL=- \, \kappa'\,e,
\ee
with
\be
\kappa'=\kappa-\frac{\gam_{\uc\ud}}2(\h^{\ua\ub}\,{\Phi_\ua}^\uc\,{\Phi_\ub}^\ud-\h^{\uc\ud}).
\label{kappad}\ee

Not all of the Euler-Lagrange (EL) equations of motion are independent but
there are $p+1$ Noether identities for diffeomorphism invariance, 
\be
\frac{\delta \CL}{\delta x^\ua}\,\partial_i x^\ua + \frac{\delta \CL}{\delta{\Phi_\ub}^{\ua}}\,\partial_i {\Phi_\ub}^{\ua} +\frac{\delta \CL}{\delta{\gamma_{\ub\ua}}}\,\partial_i \gamma_{\ub\ua}\equiv 0,\qquad (i=0,1,...,p).
\ee
The Noether identities associated with the local $SO(p,1)\times SO(D-(p+1))$ gauge invariance are\footnote{ Our convention is $\frac{\D\gam_{\ua\ub}}{\D{\gam_{\uc\ud}}}=\frac12(\D^\uc_\ua\D^\ud_\ub+\D^\uc_\ub\D^\ud_\ua)$.}
\bea
\frac{\pa\CL}{\pa  {\Phi_\uc}^{[a}}\,{\Phi_{\uc\,b]}}+2\,
\frac{\pa\CL}{\pa {\gam_\uc}^{[a}}\,\gam_{\uc\,b]}&\equiv& 0 ,
\qquad
\frac{\pa\CL}{\pa  {\Phi_\uc}^{[a'}}\,{\Phi_{\uc\,b']}}+2\,
\frac{\pa\CL}{\pa {\gam_\uc}^{[a'}}\,\gam_{\uc\,b']}\equiv 0.
\eea
They are  
$\, 
\frac{p(p-1)}{2}+ \frac{(D-(p+1))(D-p-2)}{2}$ relations among these equations of motion.

The presence of gauge symmetries and their associated Noether identities results in the appearance of the
first class constraints in the Hamiltonian formalism. In addition, since the variables ${\Phi_\ub}^{\ua}, \, {\gamma_{\ub\ua}}$  are non-dynamical (for the Lagrangian \bref{invariantdensity}),
the EL equations will produce Lagrange constraints (relations among coordinates and velocities). As we will see they will produce Hamiltonian secondary constraints and relations among the arbitrary functions
appearing in the Hamiltonian. 

Now we will make contact with ordinary NG action written in terms of the
coordinates of the p-brane. In fact, if we introduce the inverse vielbein
 ${e_b}^i$,  ${e_b}^i{e_i}^a={\D^a}_b$, the IH mechanism $L^{a'}=0$ like the particle case 
\bref{IH00}  and the orthonormality condition \bref{lorentz},  we can write ${\Phi^{\ua}}_b$ as
 \be
{\Phi^{\ua}}_b={e_b}^i\pa_ix^\ua,
\label{Phi0}
\ee
we have also 
\bea
{e_i}^b&=&
\h^{bc}{e_c}^j\ba g_{ji},
\label{eib0}
\eea 
where $ \ba g_{ij}$ is the induced metric, 
$ \ba g_{ij}=(\pa_ix_\ua\pa_jx^\ua). $
Taking the determinant of  \bref{eib0} we have
that the Lagrangian  \bref{final} becomes
\bea
 \CL
&=& -\kappa\sqrt{-\ba g}
\eea
which is the ordinary NG action.

\subsection{ WZ Lagrangian density}

Apart from the invariant Lagrangian density \bref{invariantdensity} valid for any dimension 
we can construct a WZ term for a $p$-brane
in dimensions $D=p+2$ using only the MC forms of the translations\footnote{See  \cite{Ramos:1997fa}\cite{Curtright:2010zz} for the case of the string and \cite{Bonanos:2008kr} for the 2-brane.} $L^{\ua}$. In fact 
\be
\Omega_{p+2}=\epsilon_{\ua_{0}\cdots\ua_{p+1}}L^{\ua_{0}}\wedge\cdots L^{\ua_{p+1}}= (\det\Phi)\,
\epsilon_{\ua_{0}\cdots\ua_{p+1}}dx^{\ua_{0}}\wedge\cdots dx^{\ua_{p+1}}
\ee
is a closed invariant $(p+2)$ form. Since $ \det\Phi=1$ we can write it as
\be
\Omega_{p+2}=d(\epsilon_{\ua_{0}\cdots\ua_{p+1}}x^{\ua_{0}}\wedge dx^{\ua_{1}}\wedge\cdots dx^{\ua_{p+1}})=d(L_{WZ}).
\ee
We can either  add $L_{WZ}$ on the NG Lagrangian \cite{Ramos:1997fa}   \cite{Curtright:2010zz} or 
consider the $p+1$ form $L_{WZ}$ solely as a particle model Lagrangian. 
The WZ term of the $p$-brane action represents the coupling of the $p$-brane to a (p+1) form Abelian gauge potential given
by $A_{p+1}=\epsilon_{\ua_{0}\cdots\ua_{p+1}}x^{\ua_{0}}\wedge dx^{\ua_{1}}\wedge\cdots dx^{\ua_{p+1}}$, this coupling breaks parity invariance for odd $p$.

\subsection{NLR action versus canonical action}

In this subsection we will  see how the Goldstone bosons of the broken Lorentz generators  are functions of the phase space variables 
of the $p$-brane. This result generalizes \bref{pPhipa} of the particle, sketched in the introduction to the $p$-brane case. We will see that the  NLR action \bref{invariantdensity} leads in a natural way to the canonical action of a $p$-brane. 
 The definition of all the canonical momenta from the action \bref{invariantdensity}\footnote{We consider
$\xi^0$ as the canonical time and use "dot" as $\xi^0$ derivative.  We often do not write the dependence on $\xi^i$. }
gives primary constraint equations  (See the details of the Hamiltonian formalism in the appendix B)
\bea
\phi_\ua&\equiv &p_\ua+\kappa'\, e\,{e_b}^0\,{\Phi_\ua}^b=0, 
\label{s1153.3a}\\
{\phi^\ua}_{\ub}&\equiv &{\Pi^\ua}_{\ub}=0,\label{s1153.4a}\\
{\phi_\gam^{\ua\ub}}&\equiv &p_\gam^{\ua\ub}=0.
\label{s1153.5aa}\eea
Note that the combination $e{e_b}^0$ does not depend on the velocities\footnote{
It is useful the relation
$e{e_a}^i=-\frac{1}{p!}\ep^{i i_1...i_p}\ep_{aa_1...a_p}\,{e_{i_1}}^{a_1}...\,{e_{i_p}}^{a_p}$.} $\dot x^\ua$
. 
The Hamiltonian is a sum of primary constraints,
\bea
\CH_D=\int d^{p}\xi\, \CH,\qquad 
\CH&=&
\phi_\ua\,\lam^\ua+{\phi^\ua}_{\ub}{\Lam_\ua}^{~\ub}+
\phi_\gam^{\ua\ub}\lam^\gam_{\ua\ub},
\label{Dhamiltonian}\eea
where $\lam^\ua, \,{\Lam_\ua}^{~\ub}$ and $\lam^\gam_{\ua\ub}$ are arbitrary functions of $\xi$ at this moment. Using the Hamilton's equations  
they are related to the velocities, 
\bea\label{arbitrary}
\dot x^\ua&=&\{x^\ua,\CH_D\} =\lam^\ua, 
\quad
\dot \gam_{\ua\ub}=\{\gam_{\ua\ub},\CH_D\} = \lam^\gam_{\ua\ub},\quad 
{\dot\Phi_\ua}{}^{\,\ub}=\{ {\Phi_\ua}^{\,\ub},\CH_D\} ={\Lam_\ua}^{\ub}.
\eea
The Hamilton's equations for momenta reproduce the EL equations. The consistency condition  that  
the primary constraints \bref{s1153.3a}-\bref{s1153.5aa} remain zero under time evolution gives the secondary constraints 
\be
\chi^{\ua\ub} \equiv C^{\ua\ub}= \frac{1}2(\h^{\uc\ud}\,{\Phi_\uc}^{\ua}\,{\Phi_\ud}^{\ub}-\h^{\ua\ub})=0, \label{s1153.42b}
\ee
\be
\gam_{b'c}=\gam_{b'c'}=0,\qquad 
\gamma_{ab}-\,{\kappa}{\h_{ab}}=0,
\label{gammaaab}\ee
\be\label{inversehiggsm}
{\chi_m}^{b'}\equiv \pa_mx^\ua{\Phi_\ua}^{b'}=0,\qquad (m=1,...,p)
\ee
and  conditions on $\lam^\ua$ and $\Lam$, 
\be\label{inversehiggs0}
\7\lam^{b'}\equiv \lam^\ua{\Phi_\ua}^{b'}=0,
\ee
\bea
 (e{e_b}^0\,{\Lam_\uc}^b\, +e{e_b}^m\,\pa_m{\Phi_\uc}^b\,){\Phi^\uc}_{a'}
=0,
\label{s52a3}\eea
where $e{e_b}^m,(m=1,...,p)$ are linear  functions of $\lam^\ua=\dot x^\ua$.  
Eq.\bref{s1153.42b} is the orthonormality of $\Phi$ and  Eq.\bref{gammaaab}
fixes the $\gamma_{\ua\ub}. $ Eq.\bref{inversehiggsm} and Eq.\bref{inversehiggs0}
coincide with the vanishing of the MC forms $L^{b'}$ in \bref{vielbein} associated to broken translations, 
which is known as the inverse Higgs mechanism (IH) \cite{Ivanov:1975zq}.

The stability of the secondary constraints does not produce any further constraints but only
conditions on the multipliers,
\be
\dot\chi^{\ua\ub} =\frac{1}2(\h^{\uc\ud}\,{\Lambda_\uc}^{\ua}\,{\Phi_\ud}^{\ub}+\h^{\uc\ud}\,{\Phi_\uc}^{\ua}\,
{\Lambda_\ud}^{\ub})= 0,\label{s1153.423c}\ee
\be
\dot\chi^\gam_{\ua\ub}=\lam^\gam_{\ua\ub}=0
\label{lamgan}\ee and
\be\label{inversehiggsm2}
\pa_mx^\ua{\Lam_\ua}^{b'}+\pa_m\lam^\ua{\Phi_\ua}^{b'}=0.
\ee
If we redefine the arbitrary functions ${\Lam_\ua}^{\ub}$ in terms of ${\W_\ua}^{\ub}$ by
${\Lam_\ua}^{\ub}\equiv {\Phi_\ua}^{\uc}{\W_\uc}^{\ub}, \label{ADDdefLamW}
$
the consistency condition \bref{s1153.423c} imposes antisymmetry of $\W^{\ua\ub}$,
and Eq.\bref{s52a3}-\bref{inversehiggsm2} are solved for   ${\W_b}^{b'}$  as 
\bea
{\W_{a'}}^b
&=&\left((e{e_{[d}}^m{e_{a]}}^{0}\, {\Phi_\ua}^{a}\pa_m{\Phi_\uc}^d\,{\Phi^\uc}_{a'})\frac{(e{e^{b0}})}{{\bf g}}-(\,\pa_m{\Phi_{\ua a'}}){\bf g}^{m\l}{e_\l}^b\right) {\lam^\ua},
\label{Wadb2}\eea
where ${\bf g}^{m\l}$ is the inverse of   the spatial world-sheet induced metric ${\bf g}_{m\l}={e_m}^\ua e_{\l\ua}$
and  ${\bf g}$ is the determinant of  ${\bf g}_{m\l}$. 
There remain arbitrary anti-symmetric 
Hamiltonian multipliers  $\W_{ab}$,  $\W_{a'b'}$ and 
$\7\lam^a\equiv\lam^\ua{\Phi_\ua}^a$ corresponding to
local $SO(p+1),SO(D-(p+1))$ and Diff$_{p+1}$ gauge invariances. 
\vs

The Hamiltonian is written as a linear combinations of constraints with 
the independent arbitrary multipliers as
\be
\CH=\hat\lam^\perp\,\CH_\perp+\hat\lam^m\,\CH_m
+\frac12\left(J_{ab}\,\7\W^{ba}+J_{a'b'}\,\7\W^{b'a'}\right) ,
\label{Ham3}\ee
where we have used the fact that products of two constraints vanish as strong equations.
The constraints appearing here 
are the first class combinations of the constraints\footnote{Notice that the first class constraint $\CH_\perp$ depends on the broken rotations 
${J^{ba'}}$, this is in agreement with results of \cite{hep-th/0607104} about local right translations.}
\bea
\CH_\perp&=&\frac1{2\kappa}(\h^{\ua\ub}p_\ua p_\ub+\kappa^2\,{\bf g})+
{J^{ba'}} (e{e_{[d}}^m{e_{b]}}^{0})(\pa_m{\Phi_\uc}^d\,{\Phi^\uc}_{a'}), \label{Diffgenerator1}
\\
\CH_m&=&p_\ua\,\pa_m x^\ua+ 
{\Pi^\ua}_\ub  \pa_m{\Phi_\ua}^\ub, \label{Diffgenerator2}
\\
J_{ab}&=&{\Pi^\uc}_{[a}\Phi_{\uc b]},\qquad 
\\
J_{a'b'}&=& {\Pi^\uc}_{[a'}\Phi_{\uc b']}
\label{Diffgenerator}\eea
and the arbitrary multipliers $\hat\lam^\perp, \hat\lam^m, \7\W^{ba}, \7\W^{b'a'}$  are given in \bref{lamtransv} and \bref{lamlam}.

The other independent constraints are the second class constraints that are used to reduce the phase space variables. In order to find them explicitly we  rewrite the variables ${\Phi_\ua}^\ub$ { in terms of new variables} as
\bea
{\Phi_\ua}^\ub&=&
\begin{pmatrix}
{B_{1a}}^{c} & {{\7\vp}_a}^{~c'}{B_{2c'}}^{d'} \cr 
-{\vp^d}_{a'}{B_{1d}}^{c}  & {B_{2a'}}^{d'} \,\end{pmatrix}
\begin{pmatrix}
{\phi_c}^{b}  &0\cr 0  & {\phi_{d'}}^{b'}\end{pmatrix},
\label{Phipara}\eea
where $B_1$ and $B_2$ are symmetric matrices defined by 
\be
{((B_1)^{-2})_a}^b=({\D_a}^b + {\vp_a}^{c'}{\vp^b}_{c'}), \qquad 
{((B_2)^{-2})_{a'}}^{~b'}=({\D_{a'}}^{b'} + {{\7\vp}^{c}}_{~a'}{\7\vp_c}^{~b'}). 
\ee
In terms of {these variables} the orthonormality constraints  \bref{s1153.42b} are written as 
\be
{\phi_a}^c{\phi^b}_c={(\phi\phi^T)_a}^b={\D_a}^b,\qquad 
{\phi_{c'}}^{a'}{\phi^{c'}}_{b'}={(\phi'^T\phi')^{a'}}_{b'}={\D^{a'}}_{b'},\qquad  {\7\vp_a}{}^{b'}= {\vp_a}^{b'}. 
\ee
It means  ${\phi_a}^b$  {is a group element of}   $SO(p,1)$  and    ${\phi_{a'}}^{b'}$ belongs to $SO(D-(p+1))$. Using the local invariances we can take
$ {\phi_a}^b={\D_a}^b,\,{\phi_{a'}}^{b'}={\D_{a'}}^{b'}. $
In this gauge the second class constraints are ${{\7\vp}^a}_{b'}={{\vp}^a}_{b'}$ 
and  ${\vp_a}{}^{b'}$ are expressed in terms of $p_\ua$ and $\pa_mx^\ua$ as\footnote{Note that in this gauge, Eq.\bref{Phipara} becomes the parametrization of Lorentz transformations used in \cite{Gomis:2006xw}.} 
\bea
{\vp_{a_0}}^{b'}&=&
\frac{-1}{p!\Delta}\ep^{i_0...i_p}\ep_{a_0...a_p}\pa_{i_1}x^{a_1}...\pa_{i_p}x^{a_p}\,\pa_{i_0}x^{b'}|_{\dot x^\ua\to p^\ua},
\label{sol}\\
\Delta&=&
\frac{-1}{(p+1)!}\ep^{i_0...i_p}\ep_{a_0...a_p}\,\pa_{i_0}x^{a_0}
\pa_{i_1}x^{a_1}...\pa_{i_p}x^{a_p}|_{\dot x^\ua\to p^\ua},
\label{Det}\eea
where the replacement $\pa_0x^\ua\to p^\ua$  is  to be done  in the right hand side of the previous equations.

Therefore the Goldstone bosons ${\vp}_{a}{}^{b'}$ associated to Lorentz broken rotations  are written as functions of the 
phase space variables of the $p$-brane. 
Now all second class constraints are used to reduce the phase space to that of { the space-time coordinates 
and momenta of a $p$-brane} $(x^\ua, p_\ua)$. 
The action of $p$-brane obtained from the non-linear realization in the reduced space leads
to the canonical action of a Dirac-Nambu-Goto $p$-brane in configuration space.
The canonical Lagrangian becomes 
\be
\CL=\int d^p\xi\,\left(p_\ua\dot x^\ua-(\hat\lam^\perp\,\CH_\perp^1+\hat\lam^m\,\CH_m^1)\right),
\label{Dhamiltonian4}\ee with
 \be
\CH_\perp^1=\frac1{2\kappa}(\h^{\ua\ub}p_\ua p_\ub+\kappa^2\,{\bf g}),\qquad
\CH_m^1=p_\ua\,\pa_m x^\ua. \label{Diffgenerator20}\ee
 
In summary the action \bref{invariantdensity}, which depends on the coordinates of the $p$-brane
and the Goldstone bosons associated to the broken Lorentz rotations, leads 
in a natural way to the canonical action of a $p$-brane.


\section{Corrections to the NG Lagrangian}

In section 2 we have constructed in generic dimensions an invariant action for $p$-brane in terms of the MC forms of unbroken translations. We have also constructed the  WZ Lagrangian (quasi invariant Lagrangian) in dimension $p+2$ for a $p$-brane using the MC forms of all translations.
 In this section we will see that we can use the MC forms of the broken Lorentz rotations to 
discuss possible terms of the effective action of a $p$-brane\footnote
{In \cite{Gauntlett:1989qe}  \cite{Gauntlett:1990nk} the MC forms associated to broken Lorentz rotations were used to construct relativistic (super) particle gauged fixed action in  two dimensions with extrinsic curvature.}.
 In order to construct  these terms we will need to use the world-volume geometry, in particular 
the  Hodge star operator on the MC forms. According to the dimensionality of the couplings, the candidate Lagrangians built up with the MC forms will be added as corrections
to the lowest order Lagrangian \bref{invariantdensity}. 
 As was mentioned in the introduction we make a
perturbative dynamical determination of the variables associated with the broken Lorentz rotations
to obtain {higher order derivative} 
corrections to the Nambu-Goto Lagrangian
instead of the minimal use of the IH condition\cite{Ivanov:1975zq}.
In the spirit of the effective theories,
 the coefficients of these terms will be small compared to the NG term.
After considering general properties of the $p$-brane we will consider in some detail the case of the particle and the string.

\subsection{$p$-brane }

First we note that the volume form for $p$-brane  is expressed in terms of Hodge  operator $*$ as
\be
\mu=L^0...L^{p}=\frac1{p+1}L^a*L_a=d^{p+1}\xi\,e,
\label{Volumep}\ee
where the Hodge $*$  on $q$ form is defined by $(L^{b_1}...L^{b_q})*(L_{a_1}...L_{a_q})=\mu
\D_{a_1}^{b_1}...\D_{a_q}^{b_q},\, ({\rm for}\; {b_1}<...<{b_q}, {a_1}<...<{a_q})$. The NG action {of a $p$-brane}
is proportional to the volume form \bref{Volumep}, see \bref{L0action}. 
The equations of motion for the NG  $p$-brane  are expressed as
\be
\D\Phi \to L^{a'}=0,\qquad \D x^\ua \to L^{ab'}*L_a=0. 
\label{EOMNGp}\ee
The first one is obtained by the variation with respect to $\Phi$ and is known as the IH condition \cite{Ivanov:1975zq}. 
In the following we construct a geometric  $p$-brane action using with the MC forms $L^a, L^{ab'}$ and solve the 
EOM of $\Phi$ iteratively. By eliminating $\Phi$ up to a given order a geometric effective action is obtained  
as a function of higher derivative of the world-volume coordinates $x^\ua$. 
     
As possible corrections to the NG action we here consider some invariant $p+1$ forms,
\be
R^{ab}*(L_aL_b),\quad   L^{ab'}*L_{ab'},\quad K^{a'}*K_{a'},\quad R^{ab}*R_{ab},
\label{RRKK}\ee
where
\be
R^{ab}=dL^{ab}+L^{ac}{L_c}^b=L^{ac'}{L^b}_{c'},\qquad
\ee
is the curvature two form and 
\be
K^{a'}=\frac{-1}{p!}\ep_{a_0a_1...a_p}L^{a_0a'}L^{a_1}...L^{a_p}=L^{aa'}*L_a
\ee
is essentially the trace of the extrinsic curvature expressed as a $p+1$ form 
transverse vector.  Explicitly 
  \bea
R^{ab}*(L_aL_b)&=&d^{p+1}\xi\,e{e_a}^{[i}{e_b}^{j]}
(\pa_i{\Phi_\ua}^a(\h^{\ua\ub}-{\Phi^\ua}_{c}{\Phi^{\ub c}})\pa_j{\Phi_\ub}^ b) 
\equiv  d^{p+1}\xi  \,e\CR,
 \nn\\
 R^{ab}*R_{ab}&=&\frac12d^{p+1}\xi \,e\,\CR\cdot\CR, \qquad 
K^{a'}=
- d^{p+1}\xi  \,e\,{e_b}^i\,(\pa_i\Phi^b\cdot\Phi^{a'})\equiv  -d^{p+1}\xi  \,e\CK^{a'} ,\qquad 
\nn\\
K^{a'}*K_{a'}
&=&d^{p+1}\xi \,e \CK^{a'}\CK_{a'} 
=d^{p+1}\xi \,e\, {e_a}^i {e_b}^j(\pa_i{\Phi_\ua}^a(\h^{\ua\ub}-{\Phi^\ua}_{c}{\Phi^{\ub c}})\pa_j
{\Phi_\ub}^ b) \nn\\
L^{ab'}*L_{ab'}
&=&d^{p+1}\xi \,e\, {e_a}^i {e^{aj}}(\pa_i{\Phi_\ua}^b(\h^{\ua\ub}-{\Phi^\ua}_{c}{\Phi^{\ub c}})\pa_j
{\Phi_{\ub b}}) .
\label{RRRKK2}\eea
More correctly $\CR$ and $\CK^{a'}$ become 
Riemannian and extrinsic curvatures at lowest order of $\B$ (NG $p$-brane) (see e.g. \bref{extrinsicc}). Note also that the first three quantities in \bref{RRKK} are not independent at the lowest order of string, but related by the 
Gauss-Codazzi equation, see for example \cite{Curtright:1986ed}.
In \bref{RRKK} we didn't consider to use the forms $L^{a'}$ because they vanish at lowest order 
of the  EOM for $\Phi$ \bref{EOMNGp}. They only contribute in higher order perturbations. 
On the other hand we do keep the extrinsic curvatures $K^{b'}$ which vanish when the EOM of $x^\ua$ of the NG brane is taken. We use the EOM of $\Phi$ but {\it not} the EOM of $x^\ua$ in constructing the effective action.  
 This procedure is different from \cite{Polchinski:1991ax} 
  \cite{arXiv:1111.5758} where  the extrinsic curvatures are ignored in the effective action by using the EOM of $x^\ua$'s at lowest order.  

The mass dimension of the volume form (NG term) is $[\mu]=m^{-1-p}$ and the terms in 
\bref{RRKK} have dimensions 
\be
[R^{ab}*(L_aL_b)]=m^{1-p},\quad [L^{ab'}*L_{ab'}]=m^{1-p},\quad 
[K^{a'}*K_{a'}]=m^{1-p} ,\quad [R^{ab}*R_{ab}]=m^{3-p}.
\label{RRKKdim}\ee
In the following we will examine the Lagrangian by adding a linear combination of 
these terms -we pick those that are independent at the lowest order- 
whose coefficients are scaled by the brane tension $\kappa, ([\kappa]=m^{p+1}),$  
\be
\CL=-\kappa'\,e\,+  \frac{\beta_1}{\kappa^\frac{1-p}{1+p}}\,e\CR+ \frac{\beta_3}{2\kappa^\frac{1-p}{1+p}}\,e\CK^{a'}\CK_{a'} + \frac{\beta_2}{\kappa^{\frac{3-p}{1+p}}}\,e\CR\cdot\CR,
\label{Lcorrect}\ee
where the Lagrange multiplier terms are included in $\kappa'$ \bref{kappad}\footnote{We could also have considered invariant terms with higher order derivatives.} and  the  constants $\beta$'s are dimensionless.

\subsection{Particle}
In case of relativistic bosonic particle $(p=0)$ the NG particle action 
is proportional to the invariant one form $L^0$. 
We define the Hodge operator on the world-line by, $(\tau={\xi^0})$,
\be
(L^0)^**(L_0)^*= d\tau\,e=(L^0)^*
\ee
which implies
\be
*d\tau=-\frac1{e}=-\frac1{(\dot x^\ua{\Phi_\ua}^0)}.
\ee
Since there is no two form $R^{ab}$ {on the world-line} we only consider the $\CK^2$ term in the action  \bref{Lcorrect}.
Using $K^{a'}=L^{0a'}$,
\be
K^{a'}*K_{a'}=L^{0a'}*{L^0}_{a'}=(d\tau{\Phi_{\ub 0}}{\dot\Phi^{\ub}}{}_{a'})
(-\frac{({\Phi_{\ua}}^0 \dot\Phi^{\ua a'})}{e})=-\frac{d\tau}{e}\,
{{\dot\Phi}^\ua}{}_0(\h_{\ua\ub}-\Phi_{\ua 0}{\Phi_{\ub}}^{0}){\dot\Phi}^{\ub 0}
\label{partformsA}
\ee
Eq.\bref{Lcorrect} becomes  
\be\label{effective22}
\CL=-m\{\,e+\frac{\gamma}2(\Phi_\ua^0\Phi^{\ua 0}+1)+
\frac \B{2m^2 e}\,
{{\dot\Phi}^\ua}{}_0(\h_{\ua\ub}-\Phi_{\ua 0}{\Phi_{\ub}}^{0}){\dot\Phi}^{\ub 0}\},
\ee
where $\beta$ is a dimensionless constant and $[m]=m^{1}$.
Redefining $\gamma$ in Eq.\bref{effective22} in order to absorb ${\Phi_{\ub}}^{0}{\dot\Phi}^{\ub 0}$ terms it becomes{ (in units $m=1$)}
\bea \CL&=&-e-\frac{\gam}2(\Phi^2+1)-\frac{\B}{2e}\dot\Phi^2,\qquad (\Phi)_\ua={\Phi_\ua}^0,\quad e=\dot x^\ua{\Phi_\ua}^0.
\label{Lag1}
 \eea 
the $\beta$ term is considered as a perturbation of the free particle action\footnote{ We could have also added terms with higher derivatives of $\Phi$, like for example  $\tilde D*L^{0a'}*(\tilde D*L^{0a'})$, where
$\tilde D$ is the covariant derivative with respect to rotations.}.
 In order to make contact with an effective Lagrangian written in terms of the geometric quantities of 
the world-line 
we solve the equations of motion of $\Phi$ perturbatively in $\B$, in terms of the higher velocities of the particle.

 The EL equations of $\Phi$ are 
  \bea \D\Phi&;& - \dot x^\ua-\gam\,\Phi^\ua +\B\,\frac{d}{d\tau}{(\frac{\dot\Phi^\ua}{e})}+\B\,\frac{\dot\Phi^2}{2e^2}\,\dot x^\ua=0,
\label{eqPhi}\eea 
and that of $\gam$ gives 
\bea \D\gam&;& \Phi^2+1=0. \label{eqgam} \eea 
Using the second we obtain 
\bea \Phi^\ua&=&\frac1{1+\B\, \frac{ (D\Phi)^2}2}[-Dx^\ua+ \B( \frac{(D\Phi)^2}{2}\,D x^\ua+ \,D^2\Phi^\ua)],
\label{eqPhi6} \eea 
where "derivation $D$" is defined by $D\equiv \frac1e\frac{d}{d\tau}$.
Although it is a dynamical equation it is solved for $\Phi$ iteratively in $\B$, 
\bea \Phi^\ua&=&-Dx^\ua+ \B \left( {(D\Phi)^2}\,D x^\ua+ \,D^2\Phi^\ua) \right)
-\B^2\left(\frac{(D\Phi)^2}2( {(D\Phi)^2}\,D x^\ua+ \,D^2\Phi^\ua)
\right)+O(\B^3).\nn\\
\label{eqPhi62} \eea 
Using  $ (Dx)^2=\frac{\dot x^2}{e^2}$ it follows
\bea
e^2&=&\frac{-\dot x^2}{1-\B^2 J_e^2},\qquad 
\\ J_e^2&\equiv&\frac{ ((D\Phi)^2)^2}4(1+(D x)^2)+ (D\Phi)^2(D^2\Phi\cdot Dx)+
(D^2\Phi)^2.
\eea
It tells that the correction of $e$ from $\sqrt{-\dot x^2}$ starts from an $O(\B^2)$ term
\be
e=\sqrt{-\dot x^2}\,\left(1+\frac{\B^2}2 J_e^2+O(\B^4)\right).
\label{expe}\ee
This result will be used to rewrite the derivative $D$ in terms of  proper time derivative 
$\frac{d}{ds}\equiv D_s=\frac1{\sqrt{-\dot x^2}} \frac{d}{d\tau} $ plus corrections.
Remember $(D_s x)^2=-1$.

The effective action is obtained by using \bref{expe} and \bref{eqPhi62} 
in the original action \bref{Lag1},
\bea \CL&=&-e(1+\frac{\B}{2}(D\Phi)^2).
\label{Lag3}
 \eea 
 The expression of $\Phi^\ua$ up to order $\B^2$ is given by
\be
\Phi^\ua=-U^\ua-\B \Sigma^{\ua}
+\B^2\,\left(-C^{\ua}+\frac{A^2}2\,\Sigma^{\ua}-\frac{(A\Sigma)}2\,A^{\ua}
-\frac{\Sigma^2}2\,U^{\ua}
\right)+O(\B^3),
\ee
\bea
U^{\ua}&=&D_sx^{\ua},
\nn\\
A^{\ua}&=&D_sU^{\ua}=D_s^2x^{\ua},
\nn\\
\Sigma^{\ua}&=&D_sA^\ub({\D_\ub}^{\ua}-\frac{D_sx_\ub D_sx^{\ua}}{(D_sxD_sx)})=D_s^3x^{\ua}-(D_s^2xD_s^2x)D_sx^{\ua},
\nn\\
\Xi^{\ua}&=&D_s\Sigma^{\ub}({\D_{\ub}}^{\ua}-\frac{D_sx_{\ub}D_sx^{\ua}}{(D_sxD_sx)})=D_s^4x^{\ua}-(D_s^2xD_s^2x)D_s^2x^{\ua}-3
(D_s^2xD_s^3x)D_sx^{\ua},
\nn\\
C^{\ua}&=&D_s\Xi^{\ub}({\D_{\ub}}^{\ua}-\frac{D_sx_{\ub}D_sx^{\ua}}{(D_sxD_sx)})
=D_s^5x^{\ua}-(D_s^2xD_s^2x)D_s^3x^{\ua}-5
(D_s^2xD_s^3x)D_s^2x^{\ua}\nn\\&&\hskip 25mm
-\{
3(D_s^3xD_s^3x)+4(D_s^2xD_s^4x)-(D_s^2xD_s^2x)^2\}D_sx^{\ua}.
\label{curvatures}\eea
where $U^{\ua}$ is the velocity, we have also the space-like vectors  acceleration, $A^{\ua}$, relativistic jerk, $\Sigma^{\ua}$,
snap $\Xi^{\ua}$,  crackle $C^{\ua}$ \cite{Dunajski:2008tg} \cite{Russo:2009yd}.
The following relations are verified by theses vectors
\bea
U^2&=&(D_sxD_sx)=-1,
\\
A^2&=&(D_s^2xD_s^2x)={\kappa_1^2},
\nn\\
(A\Sigma)&=&(D_s^2xD_s^3x)=\frac12\kappa_1D_s\kappa_1,
\nn\\
\Sigma^2&=&(D_s^3xD_s^3x)+(D_s^2xD_s^2x)^2=((D_s\kappa_1)^2+\kappa_1^2\kappa_2^2),
\eea
where $\kappa^1, \kappa^2$ are the first and second curvature of the worldline.
 The Lagrangian up to order $\B^3$ is 
 \be \CL=
-\sqrt{-\dot x^2}\left(1+\frac{\B}2 L^{(1)}+\frac{\B^2}2 L^{(2)}+\frac{\B^3}2 L^{(3)}\right)+O(\B^4). 
\label{Lgam32s} \ee 
Here
\bea  
L^{(1)} &=& {(D\Phi)^2}|_0=(D_s^2x)^2=A^2=\kappa_1^2,  \label{L1s}  
\\ L^{(2)} &=&-\left(( ({D_s}^2x)^2)^2 +({D_s}^3x)^2\right)
=-\Sigma^2=-\left(\kappa_1^2\kappa_2^2+(D_s\kappa_1)^2\right) ,\label{L2s}
\\ L^{(3)} &=&
\frac32(({D_s}^2x)^2)^3
+\frac{5}2  ({D_s}^2x)^2 ({D_s}^3x)^2
+12 ( D_s^3xD_s^2x)^2+ \,({D_s}^4x)^2
\nn\\&=&
\Xi^2+\frac12
\,A^2\, \Sigma^2 \,-(A\Sigma)^2.
\eea 

If we use another seed Lagrangian,
for instance by introducing a non-polynomial term,
\be
 {\cal L}=-e(1+  \frac{\tilde\B}2{\sqrt{(D\Phi)^2}})\,,
\label{Lgams} \ee
(Note that the last term of \bref{Lgams} is obtained from the MC forms in the worldline by computing the square root of a scalar:
$ L^0\wedge(\sqrt{*(L^{0a'}*L^{0a'})})=d\tau\sqrt{\dot\Phi^2}$.)
then the expression for $\Phi^\ua$ will change to
\be
\Phi^\ua= U^\ua+\frac{ \tilde\B}{2\sqrt{-\dot x^2}}\frac{ A^{\ua}}{ \sqrt{(\,A)^{2}}}
+...\,,\ee
and, accordingly, the Lagrangian in terms of the embedding coordinates will also change. Up to lowest order in $\tilde\beta$ it becomes,
\be \CL=
-\sqrt{-\dot x^2}\left(1+\frac{\tilde\B}2 \sqrt{\kappa_1^2} \right) \ee 
 
Notice that  the first correction in both cases, polynomial and non-polynomial, is a function of the first curvature of the world line. This result is general and it also holds when we consider the alternative procedure, already mentioned in the introduction, of constructing Diff invariant actions through the IH mechanism \bref{inversehiggs}.

\subsection{Particle in three dimensions }  
  In the particular case of three dimensions we can construct a pseudo invariant 
Lagrangian from a $H$ invariant two form $L^{01}\wedge L^{02}$
constructed from the MC forms of broken Lorentz rotations, we have
 \be 
 L^{01}\wedge L^{02}=d L^{12}, \qquad L^{12}=-d\tau \left(
\,{\dot\Phi}^{\ua 1}{\Phi_{\ua}}^2\right),
 \ee
 therefore we can consider an action of a particle in three dimensions 
\be\label{effective}
I=-m\int d\tau\{\,e-\frac{\gam_{\uc\ud}}2(\h^{\ua\ub}\,{\Phi_\ua}^\uc\,{\Phi_\ub}^\ud-\h^{\uc\ud})+\frac \B{m}\,\left(\,{\dot\Phi}^{\ua 1}{\Phi_{\ua}}^2\right)\}.
\ee
The EOM of $\gamma$ gives the orthonormality of $\Phi$'s
and that of $\Phi$ gives
\bea
\dot x^\ua-\gam_{0\uc}\,{\Phi^{\ua\uc}}&=&0, \qquad
\gam_{b'\uc}\,+\ep_{b'd'}\frac \B{m}\,{\dot\Phi}^{\ua d'}{\Phi_{\ua\uc}}=0.
\eea
Using the second one the first equation becomes
\bea
\dot x^\ua-\gam_{00}\,{\Phi^{\ua 0}}-\frac \B{m}\,{\Phi_{\ub 0}}
\,\ep_{c'd'}{\dot\Phi}^{\ub c'}\,{\Phi^{\ua d'}}=0.
\label{3.29}\eea
Saturating this last equation with ${\Phi_\ua}^{ 0}$ we get $e=-\gam_{00}$ and 
\be
{\Phi^{\ua 0}}=-\frac1e
\left(\dot x^\ua-\frac \B{m}\,{\Phi_{\ub 0}}
\,\ep_{c'd'}{\dot\Phi}^{\ub c'}\,{\Phi^{\ua d'}}\right).
\label{Phi22}\ee
On the other hand, saturating Eq.\bref{3.29} with ${\Phi_\ua}^{ b'}$, we obtain 
\bea
 \dot x^\ua{\Phi_\ua}^{ b'}=\frac \B{m}\,{\Phi_{\ub 0}}
{{\dot\Phi}^{\ub}}{}_{ c'}\,\ep^{c'b'}
\eea
which is the correction of the IH condition \bref{IH00}. 
Saturating Eq.\bref{Phi22} with ${\dot x_\ua}$ we can express $e$ as
\be
 e=\sqrt{-\dot x^2}\left(1+\frac {\B^2}{2m^2(-\dot x^2)}\,({\Phi_{\ua 0}}{\dot\Phi}^{\ua b'})^2\right). \label{e3D}\ee

 Using it and Eq.\bref{Phi22} we can rewrite the Lagrangian \bref{effective}. In doing so, 
since $SO(2)$ is a symmetry of Eq.\bref{effective}, one can make different choices for $\Phi^{\ua 
1},\,\Phi^{\ua 2}$, compatible with Eq.\bref{Phi22}. Each choice amounts to a complete gauge fixing of 
the 
$SO(2)$ invariance. A possible choice is 
\bea
{\Phi^{\ua 0}}&=&-D_sx^\ua-\frac \B{m}\,((D_s{\Phi_\ub}^0)
{\Phi^\ub}_{c'}\,\ep^{c'd'})\,{\Phi^\ua}_{ d'},
\label{Phi2230}\\
{\Phi^{\ua 1}}&=&\frac{D_s^2x^\ua}{\sqrt{{(D_s^2x)^2}}} 
-\frac \B{m}\,((D_s{\Phi_\ub}^0){\Phi^\ub}_{c'}\,\ep^{c'1}) {\Phi^{\ua 0}} ,
\label{Phi2231}\\
{\Phi^{\ua 2}}&=&\frac{\ep^{\ua\ub\uc}(D_s x_\ub)
{(D_s^2 x_{\uc})}}{\sqrt{(D_s^2 x)^2}} 
-\frac \B{m}\,((D_s{\Phi_\ub}^0){\Phi^\ub}_{c'}\,\ep^{c'2}) {\Phi^{\ua 0}}.
\label{Phi2232}\eea
The action \bref{effective} becomes, up to $O(\B^2)$
\bea\label{effective2}
I&=&
-m\int ds\left(1+
\frac \B{m}\,\left(
\,{D_s\Phi}^{\ua 1}{\Phi_{\ua}}^2\right)|_{NG}+\frac {\B^2}{m^2}
\,(D_s{\Phi_{\ua}}^{ 0} {\Phi}^{\ua b'})^2|_{NG})
\right)+O(\B^3),
\nn\\ \eea
where $\B^2$ term comes from both $e$ and $\frac \B{m}(\,{\dot\Phi}^{\ua 1}{\Phi_{\ua}}^2)$ terms in \bref{effective}.\footnote {Different parametrizations of $\Phi$'s connected by local SO(2) give equivalent action since $  \left(
\,{D_s\Phi}^{\ua 1}{\Phi_{\ua}}^2\right)|_{NG} $ differs by a surface term
and  $(D_s{\Phi_{\ua}}^{ 0} {\Phi}^{\ua b'})^2|_{NG}$ is SO(2) invariant.} 
 Here $A|_{NG}$ means $A$ is evaluated at $O(\B^0)$.
It is written as
\bea
I&=&-m\int ds\left(1+
\frac \B{m}\,\kappa_2+\frac {\B^2}{m^2}\,{\kappa_1}^2 \right)+O(\B^3).
\label{3Dcorrection}\eea
The $O(\B)$ term $\kappa_2$ is the torsion 
and $\kappa_1$ in $O(\B^2)$ term is the curvature, all defined in  \bref{L1s}-\bref{L2s}. 
Particle dynamics with the torsion term has been discussed in  \cite{Plyushchay:1990yf}  
and  in non-covariant gauge -corresponding to another choice instead of \bref{Phi2231},\bref{Phi2231}- 
in \cite{Schonfeld:1980kb} \cite{Mezincescu:2010gb}. Both Lagrangians differ by a total derivative. 

\subsection{String}
 
The NG action for the string is constructed from the $SO(1,1)\times SO(D-2)$ invariant two form $\frac12\ep_{ab}L^a\wedge L^b=L^0\wedge L^1\equiv \mu$, which is the volume form of the world-sheet \bref{L0action}.
The Hodge operator for one form  is
\be
L_a\wedge *L^b={\D_a}^b\,\mu,\qquad \mu=d^2\xi\,\det e
\label{4.22}\ee
then the Hodge operations on $L^a$ and $d\xi^i$ are
\be
 *L_a=-\ep_{ab}L^b, \qquad *d\xi^i =-e\,g^{ij}\ep_{jk}d\xi^k,
\label{4.23}\ee
where $g^{ij}$ is the inverse of the metric of the world-sheet $g_{ij}=e_i{}^a e_i{}^b\eta_{ab}$ and
$ e=\det e=-\frac12(\ep_{ab}\ep^{ij}{e_i}^a{e_j}^b) $.
Using \bref{RRRKK2} we can write the invariant scalar local density \bref{Lcorrect} for the string as
\be
\CL=-\kappa'\,e\,+ \frac{\beta_1}{\kappa}\,e\CR +\frac{\beta_3}{2} \,e\CK^2+ \frac{\beta_2}{\kappa}\,e\CR^2.
\label{lag-str}\ee
The dimension of the string tension is $[\kappa]=m^{2}$ and the $\B$'s are dimensionless. 
The Lagrangian is described by $ x^\ua, {\Phi_\ua}^\ub $ and the Lagrange multipliers $\gam_{\uc\ud}$.     
The first term is the NG Lagrangian and the second term is 
a total divergence. If we ignore the surface term Eq.\bref{lag-str} becomes
\bea
\CL&=& -\kappa\,e+\frac{e}2\gam_{\uc\ud}(
{\Phi_\ua}^\uc{\Phi_\ub}^\ud\h^{\ua\ub}-\h^{\uc\ud})+
\frac{\beta_2}{\kappa\,e }\,
{\left(\ep_{ab}\ep^{ij}\pa_i{\Phi_\ua}^a\,O_\Phi^{\ua\ub}\,
\pa_j{\Phi_\ub}^b\right)^2}
\nn\\&&+\frac{\beta_3}{2}{\,e }
\,{({e_a}^i\pa_i{\Phi_\ua}^a)O_\Phi^{\ua\ub}
({e_b}^j\pa_j{\Phi_\ub}^b)}
\equiv -\kappa'\,e+\B\,\CL'.
\label{Lagb2}\eea
where $O_\Phi^{\ua\ub}=(\h^{\ua\ub}-{\Phi^{\ua c}}{\Phi^\ub}_{c})$. 
For small $\B$'s we may solve the EOM for $\Phi$ perturbatively 
and the resulting {geometrical} Lagrangian depends on higher order derivatives of $x$. 
 The EOM of $\Phi$ is 
\bea
-\kappa e{e_b}^i\pa_ix^\ua{\D_\ub}^b+e\gam_{\ub\ud}\Phi^{\ua\ud}+\B
{(\CL')^\ua}_\ub=0,\qquad  {(\CL')^\ua}_\ub\equiv\frac{\D\CL'}{\D{\Phi_\ua}^\ub}.
\label{eomPhi}\eea
Since $\CL'$ does not depends on ${\Phi_\ua}^{b'}$,  
$ {(\CL')^\ua}_{b'}=0$. 
Taking $\ub=b'$ in Eq.\bref{eomPhi} we obtain
\be
\gam_{b'd}=\gam_{b'd'}=0,
\label{gam1}\ee
whereas from the $\ub=b$ component of Eq.\bref{eomPhi} we can determine
\be
\gam_{ab}=\kappa \h_{ab}-\frac{\B}{e}
{(\CL')^\ua}_a\Phi_{\ua b}, \qquad {(\CL')^\ua}_{[a}\Phi_{\ua b]}=0.
\label{gam2}\ee
The latter equality holds identically due to the local $SO(1,1)$ invariance of $\CL'$ under \\ $\D{\Phi_\ua}^b= {\A^b}_{c}{\Phi_\ua}^c,\, (\A^{bc}+\A^{cb}=0)$. 
Plugging \bref{gam1} and \bref{gam2} into \bref{eomPhi},
\be
{\Phi^{\ua}}_b={e_b}^i\pa_ix^\ua-\frac{\B}{\kappa e}O_\Phi^{\ua\ub} {(\CL')_{\ub b}}.
\label{Phi}\ee
 
Now, saturating Eq.\bref{Phi} with ${\Phi_\ua}^{c'}$ and using the orthogonality relations we get  
\be
\pa_ix^\ua{\Phi_\ua}^{c'}=\frac{\B}{\kappa e}\,{e_i}^b {{(\CL')^\ua}_b}{\Phi_\ua}^{{ c'}},
\label{IH222}\ee
which is the correction of the IH condition.  Multiplying  $\pa_ix_\ua$ on \bref{Phi}
\bea
{e_i}^b&=&\pa_ix^\ua{\Phi_{\ua}}^b=
\h^{bc}{e_c}^j\ba g_{ji}-(\frac{\B}{\kappa e})^2 {e_i}^b {(\CL')_{\ua c}}
O_\Phi^{\ua\ub} {(\CL')_{\ub}}^{{ c}},
\eea 
where $ \ba g_{ji}$ is the induced metric, 
\be \ba g_{ji}=(\pa_jx_\ua\pa_ix^\ua).
\ee
Taking its determinant we get an expansion of $e$ 
\be
e=\det e=\sqrt{-\ba g}\left(1-\frac12(\frac{\B}{\kappa e})^2 
{(\CL')_{\ua b}}
O_\Phi^{\ua\ub} {(\CL')_{\ub}}^{{ b}}\right)+O(\B^3).
\label{eee}
\ee
We use \bref{Phi} and \bref{eee} to rewrite the Lagrangian  \bref{Lagb2}  up to  $O(\B^2)$ as
\bea
 \CL
&=& -\kappa\sqrt{-\ba g}\left(1+\B\,\CL'|_{NG} +\left(\frac12(\frac{\B}{\kappa e})^2 
{(\CL')_{\ua b}}
O_\Phi^{\ua\ub} {(\CL')_{\ub}}^{{ b}}\right)|_{NG}\right)
\nn\\
&=& -\kappa\sqrt{-\ba g}\left(1+\frac{\B_2}{\kappa}\,\CR^2|_{NG} +\frac{\B_3}2\,\CK^2|_{NG}+\left(\frac12(\frac{\B}{\kappa e})^2 
{(\CL')_{\ua b}}
O_\Phi^{\ua\ub} {(\CL')_{\ub}}^{{ b}}\right)|_{NG}\right).\nn\\
\label{Lagb24}\eea
Here $A|_{NG}$ means $A$ is evaluated at $O(\B^0)$  order, 
where the IH condition $\pa_ix^\ua{\Phi_\ua}^{a'}=0$ holds and ${\Phi_\ua}^{a}$'s are tangential to the world-sheet. 

Up to $SO(1,1)$ gauge freedom we can choose, for example, 
\be
{\Phi^{\ua{0}}}=-\frac{\dot x^\ua}{\sqrt{-\dot x^2}},\qquad 
{\Phi^{\ua{1}}}=\frac{x^{'\ua}_\perp}{\sqrt{(x^{'}_\perp)^2}},\qquad x^{'\ua}_\perp=x^{'\ua}-\frac{x'\cdot\dot x}{\dot x^2}\dot x^\ua.
\ee
The  $\CR|_{NG}$  is  the scalar curvature of the world-sheet 
\be
\CR|_{NG}=\frac1{(-\ba g)}\,\ep^{i_1i_2}\ep^{j_1j_2}\,(\pa_{i_1}\pa_{j_1}x^\ua)
( \h_{\ua\ub}-\ba g^{i_3j_3}\pa_{i_3}x_\ua \pa_{j_3}x_\ub)\,(\pa_{i_2}\pa_{j_2} x^\ub)
\ee
and the $\CK^{c'}=\CK^{c'}|_{NG}$ is  the extrinsic curvature  
 written as
\be
\CK^{c'}|_{NG}=\,(\square x^\ua){n_\ua}^{c'},
\label{extrinsicc}\ee
here $\square $ is the $\ba g_{ij}$ covariant d'Alembertian
and ${{n}_\ua}^{c'}={\Phi^{\ua{c'}}}|_{NG}$  are normal unit vectors of the world-sheet. 
The $O(\B^2)$ term of the Lagrangian in \bref{Lagb24} { should give higher order curvatures of the world-sheet.}

\section{Conclusions}

In this paper we have constructed terms of the effective action of a tensionfull $p$-brane using the  non-linear realization method. 
We have considered the Goldstone bosons associated to the  broken and unbroken translations,
 coordinates of the $p$-brane, and the
Goldstone bosons associated to the broken Lorentz rotations. In order to avoid using an explicit parametrization
of the Lorentz transformations we have introduced also Goldstone bosons
associated to the unbroken rotations
$SO(1,p)\times SO(D-(p+$1)). The Goldstone bosons of unbroken generators could be eliminated using the corresponding gauge transformations of local rotations.

We have seen using the action with the lowest order of derivatives that the Goldstone bosons of the broken Lorentz rotations are non dynamical and can be expressed as 
functions of the coordinates and momenta of the $p$-brane.
We have also seen how the unperturbed action leads in a natural way to the canonical action of the NG $p$-brane.
In other words the action obtained through the non-linear realizations of space-time symmetries is an action of phase space type, this result generalizes 
analogous results for the particle \cite{Casalbuoni:2008iy} \cite{Gibbons:2009me} to the $p$-brane.

We have constructed the lower order corrections up to velocities in $x$ and $\Phi$ to the NG action. The correction terms in the spirit of the effective action are small.
 We do not use the IH constraint\cite{Ivanov:1975zq}, instead we solve
perturbatively the equations of motion for the Lorentz variables $\Phi$ which, 
upon substitution in the original Lagrangian, give 
higher order geometrical terms for the effective action.
For the case of the particle and string, 
the actions obtained in this way are written in terms of  natural geometrical objects. 

 Our method does not have the completeness of the IH method, which
gives any possible correction terms for the NG Lagrangian of a
$p$-brane. However we want to
remark that it naturally selects some specific terms within the wider
set of potential correction terms. It appears that there is a very
natural, an non-trivial, geometric interpretation of the first terms 
obtained in our approach, which may suggest that there could be
physical reasons behind the selection of 
 the seeds that give raise to these terms. At this point we
do not have a compelling argument for these choices, but we think that
they have enough interest to be considered.

This method could be also useful in cases
where the geometrical quantities are not well known, for example non-relativistic string theories, \cite{Gomis:2000bd}  
\cite{Danielsson:2000gi}  \cite{Brugues:2004an}  \cite{Gomis:2005pg} \cite{Brugues:2006yd}, Finsler type theories
\cite{Gibbons:2007iu}
and the corresponding supersymmetrization.

\vspace{1cm}

 {\bf Acknowledgements }  We acknowledge comments from Roberto Casalbuoni, Paul Townsend and Toine Van Proeyen.
 J.G. acknowledges Toine Van Proeyen for the hospitality at KU Leuven where this work was completed.
 We also acknowledge partial financial support from projects FP2010-20807-C02-01,
2009SGR502 and CPAN Consolider CSD 2007-00042.

 \appendix

\section{MC equations  of Poincare group. Geometrical Aspects }

The MC forms of the Poincare algebra \bref{MC-forms} satisfy the MC equations 
\be
dL^\ua+L^{\ua\ub}L_\ub=0,\qquad 
dL^{\ua\ub}+L^{\ua\uc}{L_\uc}^\ub=0\,,
\label{MCPoinc}\ee 
which mean that the target space is torsionless and has no curvature.

We can consider the pullback to the world-volume $\Sigma$ and 
$L^{a*}=d\xi^i\,{e_i}^a$ as $p+1$-bein of $\Sigma$
and  $L^{ab*}=d\xi^i\,{\w_i}^{ab}$ as the spin connection.
Using \bref{MCPoinc}\footnote{All expressions below are pullbacks to the world-volume. The pullback notation, $L^{a}\rightarrow L^{a*}$, etc., is omitted for simplicity.}  
\bea
dL^a+L^{ab}L_b+L^{ab'}L_{b'}=0,\too de^a+{\w^a}_b\,e^b=-L^{ab'}L_{b'}\equiv
T^a,\\ 
dL^{ab}+L^{ac}{L_c}^b+L^{ac'}{L_{c'}}^b=0\too 
d\w^{ab}+\w^{ac}{\w_c}^b=-L^{ac'}{L_{c'}}^b\equiv R^{ab}.
\eea
Here $T^a$ is torsion two form of $\Sigma$, which vanishes for the NG $p$-brane since
$L_{b'}=0$ using the EOM. 
$R^{ab}$ is the curvature two form, 
$\displaystyle R^{ab}=\frac{1}{2}d\xi^{i}d\xi^j\,{R_{ij}}^{ab}$. It is related to the scalar curvature $\CR$ as
follows. In general $p+1$ dimensions  ($\ep^{01...}=1=-\ep_{01...}$)
\bea &&\
\frac{-1}{(p-1)!}\,\ep_{a_0...a_{p}}e^{a_0}...e^{a_{p-2}}\,R^{a_{p-1}a_p}
\nn\\&=&
\frac{-1}{2(p-1)!}\,d^{p+1}\xi\,\ep_{a_0...a_{p}}\ep^{i_0...i_{p}}{e_{i_0}}^{
a_0}...{e_{i_{p-2}}}^{a_{p-2}}\,{R_{i_{p-1}i_p}}^{a_{p-1}a_p}
\nn\\&=&
\frac{1}{2}\,d^{p+1}\xi\,e\,{e_{a_{p-1}}}^{[i_{p-1}}{e_{a_{p}}}^{i_{p}]}\,{R_{i_{p
-1}i_p}}^{a_{p-1}a_{p}},\qquad \qquad e=\det({e_i}^a),
\nn\\&=&
d^{p+1}\xi\,e\,{e_{a}}^{i}{e_{b}}^{j}\,{R_{ij}}^{ab}=\,d^{p+1}\xi\,e\,\CR,\qquad \qquad 
\qquad 
\CR={e_{a}}^{i}{e_{b}}^{j}\,{R_{ij}}^{ab}.
\eea

 In the case of the string  
($p=1$), 
\bea &&
-\ep_{ab}\,R^{ab}
=-\frac12\,d^{2}\xi\,\ep_{ab}\ep^{ij}\,\,{R_{ij}}^{ab}=\frac12\,
d^{2}\xi\,e\,{e_{a}}^{[i}{e_{b}}^{j]}\,{R_{ij}}^{ab}=\,d^{2}\xi\,e\,\CR,
\eea where 
\be 
{e_{a}}^{[i}{e_{b}}^{j]}=\,-\ep_{ab}\ep^{ij}\,\det({e_{a}}^{i})=\,-\ep_{ab}\ep^{
ij}/e\,.
\ee
In addition, in this case,
\be
 R^{ab}=-L^{ac'}{L_{c'}}^b=d\w^{ab}+\w^{ac}{\w_c}^b=d\w^{ab}
\ee
because $\w^{0c}{\w_c}^1\equiv 0$ for the string. Then the curvature $R^{ab}$ is an
exact form and, as a consequence, the scalar density $e\CR$ is a surface term in 2-dimensions. 


\section{Canonical formalism}

In this Appendix we present the canonical formalism of the action \bref{invariantdensity}
 \be\label{invariantdensityap}
\CL=- \, \kappa'\,e
=-\kappa\,e+e\,\frac{\gam_{\uc\ud}}2(\h^{\ua\ub}\,{\Phi_\ua}^\uc\,{\Phi_\ub}^\ud-\h^{\uc\ud})
\ee
and show how the  NLR action leads in a natural way to the canonical action of the NG $p$-brane.  We will also show that the Goldstone bosons of the broken Lorentz generators  are functions of the phase space variables of the  $p$-brane generalizing 
the result of the particle in the introduction.

The canonical momenta are\footnote{We consider $\xi^0$ as the canonical time and use "dot" as $\xi^0$ derivative.  We often do not write the dependence on $\xi^i$. }
\bea
p_\ua&=&\frac{\pa \CL}{\pa\dot x^\ua}= -\kappa'\, e{e_b}^0\, \frac{\pa {e_0}^b}{\pa\dot x^\ua}
=- \kappa'\,e\,{e_b}^0\,{\Phi_\ua}^b,\label{ADDs1153.30a}
\\
{\Pi^\ua}_{\ub}&=&\frac{\pa \CL}{\pa{{{\dot\Phi}_\ua}{}^{\ub}}}=0,
\label{ADDs1153.50a}
\\
{p_\gam^{\ua\ub}}&=&\frac{\pa \CL}{\pa{\dot\gam_{\ua\ub}}}=0,
\label{ADDs1153.40a}
\eea
where ${e_b}^i$ is the inverse vielbein ${e_b}^i{e_i}^a={\D^a}_b$. We have $ 
e{e_a}^i=-\frac{1}{p!}\ep^{i i_1...i_p}\ep_{aa_1...a_p}\,{e_{i_1}}^{a_1}...\,{e_{i_p}}^{a_p}$.
Note that the combination  $e{e_b}^0$ does not depend on the velocities $\dot x^\ua$. Therefore all the definitions of momenta \bref{ADDs1153.30a}-\bref{ADDs1153.40a} yield primary constraints,
\bea
\phi_\ua&\equiv &p_\ua+\kappa'\, e\,{e_b}^0\,{\Phi_\ua}^b=0, 
\label{ADDs1153.3a}\\
{\phi^\ua}_{\ub}&\equiv &{\Pi^\ua}_{\ub}=0,\label{ADDs1153.4a}\\
{\phi_\gam^{ab}}&\equiv &p_\gam^{\ua\ub}=0.
\label{ADDs1153.5aa}\eea
 The Hamiltonian is a sum of primary constraints,
\bea
\CH_D=\int d^{p}\xi\, \CH,\qquad 
\CH&=&
\phi_\ua\,\lam^\ua+{\phi^\ua}_{\ub}{\Lam_\ua}^{~\ub}+
\phi_\gam^{\ua\ub}\lam^\gam_{\ua\ub},
\label{ADDDhamiltonian}\eea
where $\lam^\ua, \,{\Lam_\ua}^{~\ub}$ and $\lam^\gam_{\ua\ub}$ are arbitrary functions of $\xi$ at this moment. 

In order to compute the Hamiltonian equations of motion we 
introduce the Poisson brackets, ( $ \xi^0$ is temporal and $ \xi^m,\; m=1,...,p$, are 
 spatial world-volume coordinates) 
\bea
&&\{x^\ua(\xi^0,\xi^m), p_\ub(\xi^0,{\xi'}^m) \}=\delta^\ua_\ub\,\delta^p(\xi^m-{\xi'}^m)\\
&&\{\gamma_{\ua\ub}(\xi^0,\xi^m), p_\gamma^{\uc\ud}(\xi^0,{\xi'}^m) \}=\frac 12 (\delta_\ua^\uc\delta_\ub^\ud
+\delta_\ua^\ud\delta_\ub^\uc)\,
\delta^p(\xi^m-{\xi'}^m)
\\
&&\{ \Phi_\ua{}^{\ub}(\xi^0,\xi^m), \Pi^\uc{}_{\ud}(\xi^0,{\xi'}^m) \}=\delta_\ua^\uc\delta_\ud^\ub\,
\delta^p(\xi^m-{\xi'}^m)
\eea
from which we obtain Hamilton's equations for the configuration variables,
\bea\label{ADDarbitrary}
\dot x^\ua&=&\{x^\ua,\CH_D\} =\lam^\ua, 
\nn\\
\dot \gam_{\ua\ub}&=&\{\gam_{\ua\ub},\CH_D\} = \lam^\gam_{\ua\ub},\nn\\ 
{\dot\Phi_\ua}{}^{\,\ub}&=&\{ {\Phi_\ua}^{\,\ub},\CH_D\} ={\Lam_\ua}^{\ub},
\eea
and the momenta,
\bea
\dot p_\ua&=& \{p_a,\CH_D\} =\pa_m \left(\frac{\pa}{\pa\pa_m x^\ua} 
(\kappa'\,e\,{e_b}^0\,{\Phi_\uc}^b\lam^\uc)\right)=-\pa_m \left(\frac{\pa}{\pa\pa_m x^{\ua}} \CL\right),\nn\\
{\dot p}_\gam^{\ua\ub}&=& \{p_\gam^{\ua\ub},\CH_D\} =-\frac{e}2(\h^{\uc\ud}\,{\Phi_\uc}^{\ua}\,{\Phi_\ud}^{\ub}-\h^{\ua\ub}) ,\nn\\ 
{{\dot\Pi}^\ua}_{\;\ub}&=&\{{\Pi^\ua}_{\,\ub},\CH_D\} =- \frac{\pa}{\pa{
{\Phi_\ua}^{\,\ub} }}\left(
\kappa'\,e\,{e_b}^0\,{\Phi_\ua}^b\lam^\ua\right)
=  \frac{\pa}{\pa{\Phi_\ua}^{\,\ub}}\left(\CL\right).
\label{ADDs11pdots}
\eea
The relations \bref{ADDarbitrary} determine the velocities $\dot x^\ua, \,{\dot \Phi_\ua}^{~\ub}$ and $\dot\gam^\gam_{\ua\ub}$   in terms of  the multipliers $\lam^\ua, \,{\Lam_\ua}^{~\ub}$ and $\lam^\gam_{\ua\ub}$  respectively. 
Here and hereafter the velocities  $\dot x^\ua, \,{\dot \Phi_\ua}^{~\ub}$ and $\dot\gam^\gam_{\ua\ub}$, when appear in the Hamiltonian analysis, are to be understood as the multipliers $\lam^\ua, \,{\Lam_\ua}^{~\ub}$ and $\lam^\gam_{\ua\ub}$. 

The consistency condition of the primary constraints $\dot\phi=0$ 
reproduces the EL equations of $ \gam^{\ua\ub},{\Phi_\ua}^{\ub}$ and  $x^\ua$ , 
\bea
\dot\phi_\gam^{\ua\ub}&=&\dot p_\gam^{\ua\ub}= -\frac{e}2(\h^{\uc\ud}\,{\Phi_\uc}^{\ua}\,{\Phi_\ud}^{\ub}-\h^{\ua\ub})=\left(\frac{\pa}{\pa\gam_{\ua\ub}} \CL\right)=0,
\label{ADDs1153.43}\\
\dot{\phi^\ua}_{\ub}&=&\dot{\Pi^\ua}_{\ub}=-\frac{\pa \CH}{\pa{\Phi_\ua}^{\ub}}= \frac{\pa}{\pa{\Phi_\ua}^{\ub}}
\CL\,=0,
\label{ADDs1153.53}\\
\dot\phi_\ua&=&\dot p_\ua+ \pa_\tau(\kappa'\, e\,{e_b}^0\,{\Phi_\ua}^b)=-\pa_m \left(\frac{\pa}{\pa\pa_m x^{\ua}} \CL\right) -
\pa_0\left(\frac{\pa}{\pa\dot x^\ua} \CL\right)
=0.\label{ADDs1153.33}\eea
From Eq.\bref{ADDs1153.43} we get $\frac{D(D+1)}2$ secondary constraints,
which are the orthonormality conditions for ${\Phi_\ua}^\ub$  
\be
\chi^{\ua\ub} \equiv C^{\ua\ub}= \frac{1}2(\h^{\uc\ud}\,{\Phi_\uc}^{\ua}\,{\Phi_\ud}^{\ub}-\h^{\ua\ub})=0. \label{ADDs1153.42b}
\ee
 The $D^2$ conditions \bref{ADDs1153.53} give
\be
- \,\kappa'\,e\,\delta_{\ub}{}^b{e_b}^i\,\pa_ix^\ua+e\,\gam_{\ub\uc}\Phi^{\ua \uc}=0,
\label{ADDs11pdots2}
\ee
where $\gam_{\ub\uc}$ is symmetric by definition. 
The $\ub=b'$ components of Eq.\bref{ADDs11pdots2} are $\displaystyle(D-(p+1))(p+1)+\frac{(D-(p+1))(D-p)}2$ secondary 
constraint,
\be
\chi^\gam_{b'c}\equiv\gam_{b'c}=0,\qquad
\chi^\gam_{b'c'}\equiv\gam_{b'c'}=0 
,\label{ADDgamsec1}
\ee
whereas the $\ub=b$ components of Eq.\bref{ADDs11pdots2}
\be\label{hola}
 - \,\kappa\,e\,{e_b}^i\,\pa_ix^\ua+e\,\gam_{bc}\Phi^{\ua c}=0
\ee
imply $\displaystyle\frac{(p+1)(p+2)}2$ secondary constraints,
\be
\chi^\gam_{ab}\equiv\gamma_{ab}-\,{\kappa}{\h_{ab}}=0,
\label{ADDgammaaab}\ee
and  $(p+1)(D-(p+1))$  relations
\be\label{ADDinversehiggs}
\pa_ix^\ua{\Phi_\ua}^{b'}=0.
\ee

 Notice that Eq.\bref{ADDinversehiggs} coincides with the vanishing of the MC forms $L^{b'}$ in 
\bref{MCPoinc}, associated with broken translations. The vanishing of these forms  
is known as the inverse Higgs mechanism \cite{Ivanov:1975zq}.
 Eq.\bref{ADDinversehiggs} gives $p(D-(p+1))$ secondary constraints  for $i=m$;
\be\label{ADDinversehiggsm}
{\chi_m}^{b'}\equiv \pa_mx^\ua{\Phi_\ua}^{b'}=0,\qquad (m=1,...,p).
\ee
and  $(D-(p+1))$ conditions on the multipliers $\lam^\ua=\dot x^\ua$ for $i=0$, 
\be\label{ADDinversehiggs0}
\7\lam^{b'}\equiv \lam^\ua{\Phi_\ua}^{b'}=0.
\ee

The  $D$ equations \bref{ADDs1153.33} do not give secondary constraints, but only $(D-(p+1))$ relations among the arbitrary functions. Let us find them. Using the notational convention mentioned after Eq.\bref{ADDs11pdots}  we have
\bea
\dot\phi_\ua&=& \pa_i(\kappa\, {e_b}^i\,{\Phi_\ua}^b)
=\pa_i \left(\frac{-\kappa}{p!} \ep^{i i_1...i_p}\ep_{ba_1...a_p}\,{e_{i_1}}^{a_1}...\,{e_{i_p}}^{a_p}\,
{\Phi_\ua}^{b} \right)
\nn\\&=&
{\kappa} \,e {e_{[b}}^i\, {e_{a_1]}}^{i_1}\,
(\pa_{i_1}x^\ub\pa_i{\Phi_\ub}^{a_1})\,
{\Phi_\ua}^{b}+\kappa\,e{e_b}^i\, \pa_i {\Phi_\ua}^{b} 
\eea
 where
\be
\frac{-1}{(p-1)!} \ep^{i i_1...i_p}\ep_{ba_1...a_p}\,
{e_{i_2}}^{a_2}...\,{e_{i_p}}^{a_p}\,
={e} {e_{[b}}^i\, {e_{a_1]}}^{i_1}\,.
\ee
Then, using Eq.\bref{ADDinversehiggs} and the constraints \bref{ADDs1153.42b}, we obtain, for $\dot\phi_\ua$, 
\bea
\dot\phi_\ua
&=&
{\kappa} \,e {e_{[b}}^i\, {e_{a_1]}}^{i_1}\,
(\pa_{i_1}x^\ub {\Phi_\ub}^d )( {\Phi^\uc}_d\pa_i{\Phi_\uc}^{a_1})\,
{\Phi_\ua}^{b}+\kappa\,e{e_b}^i\, \pa_i {\Phi_\ua}^{b} 
\nn\\
&=&\kappa\, e{e_b}^i\,\pa_i{\Phi_\uc}^b\,{\Phi^\uc}_{a'}\,{\Phi_\ua}^{a'}=0,
\eea 
showing that only $D-(p+1)$ components of $\dot\phi_\ua$, which can be conveniently taken as $\dot\phi_\ua{\Phi^\ua}_{a'}$, give independent conditions,
\bea
 (e{e_b}^0\,{\Lam_\uc}^b\, +e{e_b}^m\,\pa_m{\Phi_\uc}^b\,){\Phi^\uc}_{a'}
=0,
\label{ADDs52a3}\eea
where $e{e_b}^m,(m=1,...,p)$ are linear  functions of $\lam^\ua=\dot x^\ua$.  

\vspace{4mm}

We should further examine the stability of the secondary constraints.
 Let us first consider the conditions associated with the constraints 
$\chi^{\ua\ub}$  in \bref{ADDs1153.42b}, 
\be
\dot\chi^{\ua\ub} =\frac{1}2(\h^{\uc\ud}\,{\Lambda_\uc}^{\ua}\,{\Phi_\ud}^{\ub}+\h^{\uc\ud}\,{\Phi_\uc}^{\ua}\,
{\Lambda_\ud}^{\ub})= 0.\label{ADDs1153.423c}\ee
If we redefine the arbitrary functions ${\Lam_\ua}^{\ub}$ in terms of ${\W_\ua}^{\ub}$ by
\be
{\Lam_\ua}^{\ub}\equiv {\Phi_\ua}^{\uc}{\W_\uc}^{\ub}, \label{ADDdefLamW2}
\ee
then the consistency condition \bref{ADDs1153.423c} imposes antisymmetry of $\W^{\ua\ub}$,
 \be
\dot\chi^{\ua\ub} =\frac{1}2(\W^{\ua\ub}+\W^{\ub\ua})=0.
\label{ADDs53d}\ee
On the other hand, the consistency condition of $\chi^\gam_{\ua\ub}=0$ determines the $\displaystyle\frac{D(D+1)}2$ 
multipliers  ${\lam^\gam_{\ua\ub}}$,
\be
\dot\chi^\gam_{\ua\ub}=\lam^\gam_{\ua\ub}=0.
\label{ADDlamgan}\ee
Finally, the consistency of ${\chi_m}^{b'}=0$ in \bref{ADDinversehiggsm} is
\be\label{ADDinversehiggsm2}
\pa_mx^\ua{\Lam_\ua}^{b'}+\pa_m\lam^\ua{\Phi_\ua}^{b'}=
{e_m}^b{\W_b}^{b'}-\lam^\ua\,\pa_m{\Phi_\ua}^{b'}=0,
\ee
which are $p(R-p-1)$ linear relations among multipliers ${\W_b}^{b'}$ and 
$\lam^\ua$. 

This finishes the analysis of constraints. No tertiary constraints arise because the dynamical consequences of the secondary constraints boil down to the partial determination of the arbitrary multipliers. This is a consistent dynamical system.

\vspace{4mm}

 Let us examine for further use this partial determination of the arbitrary multipliers. Combining \bref{ADDinversehiggsm2} with \bref{ADDs52a3},
we obtain  $(p+1)(D-(p+1))$  equations
\be \label{inversehiggsm22} 
e{e_b}^0\,{\W_{a'}}^b\,+e{e_b}^m\,\pa_m{\Phi_\uc}^b\,{\Phi^\uc}_{a'}
=0,\qquad {e_m}^b{\W_b}^{b'}-\lam^\ua\,\pa_m{\Phi_\ua}^{b'}=0
\ee
that can be solved for  $(p+1)(D-(p+1))$ components
${\W_{ba'}}=-{\W_{a'b}}$ in terms of $\lam^\ua$. 
To do it we use the fact that  
\be
(e{e_a}^0, {e_m}^a), \quad (a=0,1,...,p)
\label{thebasis}
\ee 
is a complete set basis in terms of canonical variables, that is, for any $A_a$ we can make the decomposition
\be
A_a=(e{e_a}^0)\7a_0+{e_{ma}}\7a^m,\quad \Leftrightarrow  \quad 
\7a_0=e{e_a}^0\h^{ab}A_b/(e^2g^{00}),\quad \7a^m={\bf g}^{m\l}{e_\l}^bA_b ,
\ee
where ${\bf g}^{m\l}$ is the inverse of the $p\times p$ matrix of  $g_{m\l}={e_m}^a{e_\l}^b\h_{ab}
$, $\;{ g}^{00}={e_a}^0{e_b}^0\h^{ab}=\frac{\det g_{\l m}}{g}=\frac{{\bf g}}{-e^2},\; g=\det g_{ij}$ and ${\bf g}=\det g_{m\l}$. It follows the completeness relation
\bea
{\D_a}^b&=& 
-\frac{(e{e_a}^0)(e{e^{b0}})}{{\bf g}}+{e_{ma}}{\bf g}^{m\l}{e_\l}^b. 
\label{compl}\eea
Using it ${\W_{a'}}^b$ is expressed in terms of $\lam^{\ua}$ as 
\bea
{\W_{a'}}^b&=&{\W_{a'}}^a\left(-\frac{(e{e_a}^0)(e{e^{b0}})}{{\bf g}}
+{e_{ma}}{\bf g}^{m\l}{e_\l}^b\right)
\nn\\
&=&(e{e_d}^m\,\pa_m{\Phi_\uc}^d\,{\Phi^\uc}_{a'})\frac{(e{e^{b0}})}{{\bf g}}
+(-\lam^\ua\,\pa_m{\Phi_{\ua a'}}){\bf g}^{m\l}{e_\l}^b.\label{Wadb}\eea
Let us remind that only $p+1$ 
components of  $\lam^\ua$ are independent, as expressed in \bref{ADDinversehiggs0},
and that the anti-symmetric parts of ${\W_{ab}}$ and ${\W_{a'b'}}$ remain undetermined.

\vspace{4mm}

In summary we have the primary constraints  (with their number in parenthesis)
\bea
\phi_\ua&=&p_\ua+\kappa'\, e\,{e_b}^0\,{\Phi_\ua}^b=0, \qquad (D) ,
\label{ADDs1153a.3S}\\
{\phi^\ua}_{\ub}&=&{\Pi^\ua}_{\ub}=0, \qquad\qquad\qquad \quad  
\;(D^2) ,\label{ADDs1153a.4S}\\
{\phi_\gam^{\ua\ub}}&=&p_\gam^{\ua\ub}=0, \qquad\qquad\qquad   (\frac{D(D+1)}2) ,
\label{ADDs1153a.5S}\eea
and the secondary constraints 
\bea
\chi^{\ua\ub} &=&\frac{1}2(\h^{\uc\ud}\,{\Phi_\uc}^{\ua}\,{\Phi_\ud}^{\ub}-\h^{\ua\ub})
=0,\qquad (\frac{(D(D+1)}2),\label{ADDs1153.42S}
\\
\chi^\gam_{ab}&=&\gamma_{ab}-\,{\kappa}{\h_{ab}}=0,\qquad\qquad\qquad (\frac{(p+1)(p+2)}2),
\label{ADDgammaabSa}\\
\chi^\gam_{a'b}&=&\gamma_{a'b}=0, \qquad\qquad\qquad\qquad(p+1)(D-(p+1)),\label{ADDsegona}\\
\chi^\gam_{a'b'}&=&\gamma_{a'b'}=0, \qquad\qquad\qquad\qquad(\frac{(D-(p+1))(D-p)}{2}),\label{ADDsegona1}\\
{\chi_m}^{b'}&=&\pa_m x^\ua{\Phi_\ua}^{b'}=0,\qquad\qquad\qquad\quad (p(D-(p+1))) . \label{ADDs11pdots5S}
\eea

The conditions for the Hamiltonian multipliers are 
Eq.\bref{ADDlamgan} for $\lam^\gam_{\ua\ub}$
\be
\lam^\gam_{\ua\ub}=0,\qquad (\frac{D(D+1)}2),\label{ADDs53d22}\ee
Eq.\bref{ADDs53d} for $\Lam=\Phi\W$
\be
\W^{\ua\ub}+\W^{\ub\ua}=0, \qquad (\frac{D(D+1)}2),\label{ADDs53d23}\ee
and Eq.\bref{ADDinversehiggs0} for $\lam^\ua$
\be\label{ADDinversehiggs02}
\7\lam^{b'}=\lam^\ua{\Phi_\ua}^{b'}=0,\qquad (D-(p+1)).
\ee
In addition there are $(p+1)(D-(p+1))$ linear relations \bref{ADDs52a3}+\bref{ADDinversehiggsm2}
for $\lam^\ua$ and  ${\W_b}^{b'}$ which are solved for   ${\W_b}^{b'}$  as  Eq.\bref{Wadb},
\bea
{\W_{a'}}^b&=&\left((e{e_{[d}}^m{e_{a]}}^{0}\, {\Phi_\ua}^{a}\pa_m{\Phi_\uc}^d\,{\Phi^\uc}_{a'})\frac{(e{e^{b0}})}{{\bf g}}
-(\,\pa_m{\Phi_{\ua a'}}){\bf g}^{m\l}{e_\l}^b\right) {\lam^\ua}.
\label{ADDWadb2}\eea
The remaining arbitrary  
Hamiltonian multipliers are  $\W_{ab}$ (in number $\displaystyle\frac{p(p-1)}2$), 
 $\W_{a'b'}$ (in number $\displaystyle\frac{(D-(p+1))(D-p-2)}2$), and  $\7\lam^a\equiv\lam^\ua{\Phi_\ua}^a$  
(in number $\displaystyle p+1$). As expected, 
they correspond respectively to the local $SO(p+1),\,SO(D-(p+1))$ and Diff$_{p+1}$ gauge invariances. 
\vs

\subsection{First and Second class constraints}
We classify the constraints into the second class constraints that reduce the 
dependent canonical variables and the first class constraints that generate 
the gauge transformations. 
The constraints  \bref{ADDs1153a.5S} and \bref{ADDgammaabSa}-\bref{ADDsegona1} are  the  second class constraints that allow us to 
reduce the canonical pairs $(\gam,p_\gam)$,
\be
p_\gam^{\ua\ub}=0, \qquad \gamma_{ab}=\,{\kappa}{\h_{ab}},\qquad \gamma_{ab'}=\gamma_{a'b'}=0.
\ee

The $D^2$ constraints \bref{ADDs1153a.4S} are rearranged as\footnote{
Our convention of (anti-)symmetrizations are $A_{[a}B_{b]}=A_aB_b-A_bB_a$ and   $A_{(a}B_{b)}=A_aB_b+A_bB_a.$}
\be
K_{\uc\ud}\equiv {\Pi^\ua}_{(\uc} {\Phi_\ua}_{\ud)}=0,\,\qquad
J_{cd'}\equiv {\Pi^\ua} _{[c}{\Phi_\ua}_{d']}=0,
\label{ADDKJ2cl}\ee
and 
\be
J_{cd}\equiv {\Pi^\ua} _{[c}{\Phi_\ua}_{d]}=0,\,\qquad\qquad
J_{c'd'}\equiv {\Pi^\ua} _{[c'}{\Phi_\ua}_{d']}=0.
\label{ADDJJ1cl}\ee
The $D$ constraints $\phi_\ua$ in \bref{ADDs1153a.3S} are projected into 
\bea
\7\phi_{b}&\equiv&\phi_\ua {\Phi^\ua}_{b}=(p_\ua{\Phi^\ua}_b+\kappa\,e{e_b}^0)=0, \qquad (p+1),
\label{ADDtphibdA}\\
\7\phi_{b'}&\equiv&\phi_\ua {\Phi^\ua}_{b'}=p_\ua {\Phi^\ua}_{b'}=0
, \qquad \qquad\qquad (D-(p+1)). 
\label{ADDtphibdB}
\eea
The symmetric constraints $K_{\ua\ub}$'s, in number 
$\frac{D(D+1)}{2}$,  are combined
 with 
the same number of $ \chi^{\ua\ub} $ to form  $\frac{D(D+1)}{2}$  pairs of  second class constraints,
\be
\{ \chi^{\ua\ub},  K_{\uc\ud}\}= {\D^\ua}_{(\uc}\,{\D^\ub}_{\ud)}.
\label{ADDmoresecclass}
\ee
The ${(p+1)(D-(p+1))}$  constraints $J_{ab'}$ in \bref{ADDKJ2cl}  are paired with $\;{p(D-(p+1))}\;$  ${\chi_m}^{a'}$ in \bref{ADDs11pdots5S}
and  ${(D-(p+1))}\;$  $\7\phi_{b'}$ in \bref{ADDtphibdB}.
They are actually 
second class pairs since they satisfy non singular set of Poisson brackets; 
\bea
\left\{J_{cd'}, \7\phi^{b'}\right\}&=&
={\D^{b'}}_{d'}\,p^\ub{\Phi_{\ub c}}
\,=-\kappa\, e{e_c}^0\, {\D^{b'}}_{d'},
\nn\\
\left\{J_{cd'}, {\chi_m}^{b'}\right\}&=&
{\D^{b'}}_{d'}\,\pa_mx^\ub{\Phi_{\ub c}}={e_{mc}}\, {\D^{b'}}_{d'}.
\eea

The constraints appearing in the  Hamiltonian \bref{ADDDhamiltonian} 
with arbitrary multipliers are the first class constraints.  
 Let us start with the original Hamiltonian
\be
\CH= 
\phi_\ua\,\lam^\ua+{\phi^\ua}_{\ub}{\Lam_\ua}^{~\ub}+
\phi_\gam^{\ua\ub}\lam^\gam_{\ua\ub}
\label{ADDHam3}\ee
and obtain the primary first class constraints 
present in this Hamiltonian \bref{ADDHam3} as the combinations of primary constraints that still keep 
arbitrary multipliers attached to them. 
In doing so we use the conditions of the multipliers  \bref{ADDs53d22}-\bref{ADDWadb2}
obtained from the consistency conditions of the constraints, 
\be
\W^T+\W=0,\qquad  \lam^\gam_{\ua\ub}=0,\qquad 
\lam^\ua{\Phi_\ua}^{b'}=0,
\label{eq2}\ee
and \bref{ADDWadb2}.
In addition to above conditions on the multipliers we repeatedly use the fact that 
product of two constraints vanish as strong equation in the Hamiltonian\cite{Dirac:1950pj} . 

First we note using the completeness \bref{compl} 
\bea
\7\lam^a&=&\lam^\ua{\Phi_\ua}^a= \left(\frac{\h^{ac}(e{e_c}^0)(e{e_b}^0)}{-{\bf{ g}}}+
{e_m}^a{\bf g}^{m\l}{e_{\l b}}\right) \7\lam^b= 
-\h^{ac}(e{e_c}^0)\,\hat\lam^\perp+{e_m}^a\,\hat\lam^m, \nn\\
\label{lamperp}\eea
where $\hat\lam^\perp$ and  $\hat\lam^m$ are introduced by 
\be
\hat\lam^\perp= \frac{(e{e_b}^0)}{{{\bf g} }}{\7\lam}^b 
,  \qquad \hat\lam^m={\bf g}^{m\l}{e_{\l b}}{\7\lam}^b . \label{lamtransv}\ee
The first term of \bref{ADDDhamiltonian} $\phi_\ua\,\lam^\ua$ 
is written as 
\bea
\phi_\ua\,\lam^\ua&=&
(\phi_\ua {\Phi^\ua}_b)( \lam^\uc{\Phi_\uc}^b )=
\7\phi_b \,\7\lam^b,\qquad \7\lam^b= \lam^\uc{\Phi_\uc}^b, 
\quad
\7\phi_b =\phi_\ua {\Phi^\ua}_b. 
\label{Ham2}\eea
Using the completeness \bref{compl} 
\bea 
\7\phi_b  \7\lam^b&=&(p_\ua{\Phi^\ua}_a+\kappa\,e{e_a}^0) 
\left(\frac{\h^{ac}(e{e_c}^0)(e{e_b}^0)}{-{\bf{ g}}}+
{e_m}^a{\bf g}^{m\l}{e_{\l b}}\right)   \7\lam^b
\nn\\ 
&=&(p_\ua{\Phi^\ua}_a+\kappa\,e{e_a}^0) 
\frac{\h^{ac}(e{e_c}^0)(e{e_b}^0)}{-{\bf{ g}}} \7\lam^b+
(p_\ua{\Phi^\ua}_a+\kappa\,e{e_a}^0) {e_m}^a{\bf g}^{m\l}{e_{\l b}}  \7\lam^b
\nn\\ 
&=&-(p_\ua{\Phi^\ua}_a+\kappa\,e{e_a}^0) 
\h^{ac}\frac{\left((p_\ub{\Phi^\ub}_c+\kappa\,e{e_c}^0) -(p_\ub{\Phi^\ub}_c-\kappa\,e{e_c}^0) \right)}{2\kappa} \hat\lam^\perp 
\nn\\&&+
\phi_\ua{\Phi^\ua}_a {\pa_mx^\ub}{\Phi_\ub}^a{\bf g}^{m\l}{e_{\l b}}  \7\lam^b.
\label{compl2}\eea
The last term becomes  
\bea
\phi_\ua {\Phi^\ua}_a {\pa_mx^\ub} {\Phi_\ub}^a \hat\lam^m&=&
\phi_\ua ({\D^\ua}_\ub-{\Phi^\ua}_{a'} {\Phi_\ub}^{a'} ) {\pa_m x^\ub}\hat\lam^m
=\phi_\ua {\pa_m x^\ua} \hat\lam^m
\nn\\&=&
 (p_\ua+\kappa\,e{e_c}^0{\Phi_\ua}^c)  {\pa_m x^\ua}\hat\lam^m
=p_\ua\, {\pa_m x^\ua}\hat\lam^m
\equiv \CH_m^1\hat\lam^m.
\label{quadrham}
\eea
where the square of constraint term $(\phi_\ua{\Phi^\ua}_{a'})({\pa_mx^\ub} {\Phi_\ub}^{a'} )$ has been dropped from second to third equation\footnote{In \bref{quadrham} use has been made also of the secondary constraints \bref{ADDs1153.42S}. One can check that their contribution cancels out in the final expressions.}. 
In the last second line of \bref{compl2} we omit the square of constraint ${\phi_\ua}^2$ term and 
\bea 
&&\frac1{2\kappa}(p_\ua{\Phi^\ua}_a+\kappa\,e{e_a}^0) 
\h^{ac}(p_\ub{\Phi^\ub}_c-\kappa\,e{e_c}^0)
=\frac1{2\kappa}(p_\ua(\h^{\ua\ub}-{\Phi^\ua}_{a'}{\Phi}^{\ub a'})p_\ub
-\kappa^2\,e^2{g^{00}}) 
\nn\\
&=&\frac1{2\kappa}(p_\ua\h^{\ua\ub}p_\ub+\kappa^2\,{\bf g}) 
\equiv\,\CH_\perp^1 ,\qquad  
g_{mn}=\h_{\ua\ub}\pa_mx^\ua \pa_nx^\ub, 
\label{compl3}\eea
where again square of constraint terms $(p_\ua {\Phi^\ua}_{a'})({\Phi_\ub}^{a'}p^\ub)=
(\phi_\ua {\Phi^\ua}_{a'})({\Phi_\ub}^{a'}\phi^\ub)
$ have also been dropped from the second to the third equation. 
We arrive at
\be
\phi_\ua\,\lam^\ua=\hat\lam^\perp\,\CH^1_\perp+\hat\lam^m\,\CH_m^1,
\label{Ham22}\ee
where 
\bea
\CH_\perp^1&=&\frac1{2\kappa}(\h^{\ua\ub}p_\ua p_\ub+\kappa^2\,{\bf g}),
\qquad 
\CH_m^1=p_\ua\,\pa_m x^\ua.
\eea
\vs
The second term of \bref{ADDDhamiltonian} is
\bea
{\phi^\ua}_{\ub}{\Lam_\ua}^{~\ub}&=&{\Pi^\ua}_{\ub}{\Phi_{\ua\uc}}{\W^{\uc\ub}}=
\frac12{J_{ab}}{\W^{ba}}+\frac12{J_{a'b'}}{\W^{b'a'}}+{J_b}^{a'}{\W_{a'}}^b,
\eea
where we have used the anti-symmetry of $\W$. Note that ${\W^{ba}}$ and ${\W^{b'a'}}$ are arbitrary while ${\W_{a'}}^b$ is given in \bref{ADDWadb2}. Let us write ${J_b}^{a'}{\W_{a'}}^b$
as a linear combination of 
$\hat\lam^\perp$ and $\hat\lam^m$. To do this we use \bref{lamperp} and 
\bea
e{e_d}^m&=&
\frac{-1}{(p-1)!}\ep_{da_1...a_p}\ep^{m0m_2...m_p}{e_{0}}^{a_1}{e_{m_2}}^{a_2}... {e_{m_p}}^{a_p}
=e{e_{[d}}^m{e_{a_0]}}^{0}\, {\lam^\ua}{\Phi_\ua}^{a_0}
\nn\\&=&-e{e_{[d}}^m{e_{a]}}^{0}\,(e{e^{a0}})
\,\hat\lam^\perp+e{e_{[d}}^m{e_{a_0]}}^{0}\, {\Phi_\ua}^{a_0}(\pa_\l x^\ua\,\hat\lam^\l),
\label{Weedm}\eea
and take into account that in \bref{ADDWadb2} ${\W_{a'}}^b$ is expressed in terms of  $\hat\lam^\perp$ and $\hat\lam^m$. 
After some work we arrive at
\bea
{J_b}^{a'}{\W_{a'}}^b&=&
\left({\Pi^\ua}_\ub  \pa_m{\Phi_\ua}^\ub\, 
  +\frac12 J_{ab}({\Phi^{\uc a}}\pa_m{\Phi_\uc}^b) 
+\frac12J^{a'b'} ( {\Phi^\uc}_{a'}
\,\pa_m{\Phi^\uc}_{b'})\right) \hat\lam^m 
\nn\\
&+&
\hat\lam^\perp{J^{ba'}} (e{e_{[d}}^m{e_{b]}}^{0})(\pa_m{\Phi_\uc}^d\,{\Phi^\uc}_{a'})
.\eea
In summary the Hamiltonian is written as 
\be
\CH=\hat\lam^\perp\,\CH_\perp+\hat\lam^m\,\CH_m
+\frac12J_{ab}\,\7\W^{ba}+\frac12J_{a'b'}\,\7\W^{b'a'}
\label{Ham3A}\ee
with the first class constraints
\bea
\CH_\perp&=&\frac1{2\kappa}(\h^{\ua\ub}p_\ua p_\ub+\kappa^2\,{\bf g})+
{J^{ba'}} (e{e_{[d}}^m{e_{b]}}^{0})(\pa_m{\Phi_\uc}^d\,{\Phi^\uc}_{a'}), 
\nn\\
\CH_m&=&p_\ua\,\pa_m x^\ua+ 
{\Pi^\ua}_\ub  \pa_m{\Phi_\ua}^\ub\, 
\nn\\
J_{ab}&=&{\Pi^\uc}_{[a}\Phi_{\uc b]},\qquad 
\nn\\
J_{a'b'}&=& {\Pi^\uc}_{[a'}\Phi_{\uc b']}
\label{DiffgeneratorA}\eea and with a redefinition of arbitrary functions,
\bea
\7\W^{ba}&=&\W^{ba} +({\Phi^{\uc a}}\pa_m{\Phi_\uc}^b)  \hat\lam^m, 
\qquad \7\W^{b'a'}=\W^{b'a'}+ ( {\Phi^{\uc a'}}
\,\pa_m{\Phi_\uc}^{b'}) \hat\lam^m.\label{lamlam}
\eea
\vs

The $\frac{(p+1)p}{2}$ constraints $J_{ab}$ and  the $\frac{(D-(p+1))(D-p-2)}{2}$ constraints $J_{a'b'}$ are respectively the $SO(1,p)$ and $SO(D-(p+1))$ local generators for the $\Phi$ variables, 
 satisfying  the algebras, 
\bea
\{J_{a b }(\xi),J_{c d }(\xi')\}&=&\left(\h_{b [c }J_{a d ]}(\xi)-\h_{a [c }J_{b d ]}(\xi)\right)\D^p(\xi-\xi'),\qquad \nn\\
\{J_{a'b'}(\xi),J_{c'd'}(\xi')\}&=&\left(\h_{b'[c'}J_{a'd']}(\xi)-\h_{a'[c'}J_{b'd']}(\xi)\right)\D^p(\xi-\xi').
\eea
The Diff$_{p+1}$ generators  $\CH_\perp$ and  $\CH_m$\footnote{Regarding $p$-dimensional world-sheet integrations, $\CH_\perp$ in (\ref{DiffgeneratorA}) is a density of weight $2$ and the multiplier $\hat\lam^\perp$ is a density of weight $-1$, $\CH_m$ is a vector density of weight $1$ and the multipliers $\hat\lam^m$ behave as a vector. All this guarantees that \bref{Ham3A}  is a density of weight $1$ as expected.} are multiplied by local functions as
\be
 \CH_\perp[h]=\int d^p\xi\,h(\xi)\CH_\perp(\xi), \qquad  
\CH_m[f^m]=\int d^p\xi\,f^m(\xi)\CH_m(\xi),
\ee
and satisfy a closed algebra as
\bea
\left\{\CH_\perp[h],\CH_m[f^m]\right\}&=&\CH_\perp[h\pa_mf^m-f^m\pa_mh],\nn\\ 
\left\{\CH_\l[h^\l],\CH_m[f^m]\right\}&=&\CH_\l[h^m\pa_mf^\l-f^m\pa_mh^\l],
\nn\\
\left\{\CH_\perp[f],\CH_\perp[h]\right\}&=&
\int d\s\,
{\bf gg}^{\l m}{(f\pa_mh-h\pa_mf)} \nn\\&\times& \left( {\CH_\l}-\frac12
(J^{ab}\,({\Phi^\ub}_{b}\pa_\l\Phi_{\ub a})+
J^{a'b'}\,(\Phi_{\ub b'}\,{\pa_\l\Phi^{\ub}}_{a'}))\right).
\label{ADDDiffALge}\eea
 
The last equation means that 
the commutators of two local Diff transformations generated by $\CH_\perp$ results in a local transverse Diff $\CH_m$ and associated local Lorentz transformations $SO(1,p)$ and $SO(D-(p+1))$.  
Finally  
\bea
\left\{\CH_m[f^m],J_{ab}(\xi)\right\}&=&-\pa_m\left(f^mJ_{ab}(\xi)\right),\qquad
\left\{\CH_m[f^m],J_{a'b'}(\xi)\right\}=-\pa_m\left(f^mJ_{a'b'}(\xi)\right),\nn\\ 
\left\{\CH_\perp[h],J_{ab}(\xi)\right\}&=&\left\{\CH_\perp[h],J_{a'b'}(\xi)\right\}=0
\label{ADDDiffALgeJ}\eea
showing that $(\CH_\perp,\CH_m,J_{ab},J_{a'b'})$ forms a first class constraint set.

\vspace{3mm}

Let us consider the example of the string, where we have $a,b=0,1,\, m=\l=1$. The arbitrary functions $\hat\lam^\perp$  
and $\hat\lam^m,(m=1)$ have the familiar forms,
\be
\hat\lam^\perp= \frac{e }{{{\bf g} }}= \frac{\sqrt{-g}}{{g_{11}}}, 
\qquad \7\lam^1={\bf g}^{11}{g_{10}}=\frac{g_{10}}{g_{11}}.\ee
 $\CH_\perp$ becomes ("prime" means $\xi^1$ derivative)
 \bea
\CH_\perp &=&\frac1{2\kappa}(\h^{\ua\ub}p_\ua p_\ub+\kappa^2\,{x_\ua^{'2}})+
{J^{ba'}} \ep_{db} ({\Phi'_\uc}^d\,{\Phi^\uc}_{a'}), 
\eea
where $ (e{e_{[d}}^1{e_{b]}}^{0})=\ep_{db}$ and the second term commutes with the first term. 
One can easily check the closure of the algebra for the string case,   
\bea
\{{\CH_\perp}[f],{\CH_\perp}[h]\}={\CH_1}[fh'-f'h]-\frac12\int d\s\, (fh'-f'h)
\left(J^{ab}\,({\Phi^\ub}_{b}\Phi'_{\ub a})+
J^{a'b'}\,(\Phi_{\ub b'}\,{\Phi^{'\ub}}_{a'})\right),
\nn \eea  
\be
\{{\CH_1}[f],{\CH_\perp}[h]\}={\CH_\perp}[fh'-f'h]
,\qquad \{{\CH_1}[f],{\CH_1}[h]\}={\CH_1}[{fh'-f'h}].
\ee

\subsection{Reduced phase space and Gauge fixing}

The second class constraints are used to reduce the phase space variables. In order to find them explicitly 
it is convenient to rewrite the $D^2$ variables ${\Phi_\ua}^\ub$ 
in terms of  same numbers of coordinates 
\bea
{\phi_a}^{b};\quad  {(p+1)}^2,\quad && {\vp_a}^{b'};\quad (p+1)(D-(p+1)), \nn\\ 
{{\7\vp}_a}^{~b'};\quad (p+1)(D-(p+1)),\quad && {\phi_{a'}}^{b'};\quad (D-(p+1))^2 .
\label{ADDnewphi}\eea
as
\bea
{\Phi_\ua}^\ub&=&
\begin{pmatrix}{B_{1a}}^{c} & {{\7\vp}_a}^{~c'}{B_{2c'}}^{d'} \cr 
-{\vp^d}_{a'}{B_{1d}}^{c}  & {B_{2a'}}^{d'} \end{pmatrix}\,\begin{pmatrix}
{\phi_c}^{b}  &0\cr 0  & {\phi_{d'}}^{b'} \end{pmatrix},
\label{ADDPhipara}\eea
where $B_1$ and $B_2$ are symmetric matrices defined by 
\be
{((B_1)^{-2})_a}^b=({\D_a}^b + {\vp_a}^{c'}{\vp^b}_{c'}), \qquad 
{((B_2)^{-2})_{a'}}^{~b'}=({\D_{a'}}^{b'} + {{\7\vp}^{c}}_{~a'}{\7\vp_c}^{~b'}). 
\ee
In terms of new variables \bref{ADDnewphi}
the orthonormality constraints  \bref{ADDs1153.42S} are written as 
\be
{\phi_a}^c{\phi^b}_c={(\phi\phi^T)_a}^b={\D_a}^b,\qquad 
{\phi_{c'}}^{a'}{\phi^{c'}}_{b'}={(\phi'^T\phi')^{a'}}_{b'}={\D^{a'}}_{b'},\qquad  {\7\vp_a}^{\ b'}= {\vp_a}^{b'}. 
\ee
It means that ${\phi_a}^b$ is an element of $SO(p,1)$ and ${\phi_{a'}}^{b'}$ is an element of $SO(D-(p+1))$. 

One can adapt the canonical variables to the redefinition \bref{ADDPhipara} of the $\Phi$ configuration variables. Indeed, making a canonical transformation $\Phi\to (\phi,\phi',\vp,\7\vp)$ generated by
\be
W(\Pi,\phi,\phi',\vp,\7\vp)={\Pi^\ua}_\ub\,{\Phi_\ua}^\ub(\phi,\phi',\vp,\7\vp),
\ee
we  obtain the new momenta as $\pi_\phi=\frac{\pa W}{\pa\phi}$, etc. 
The primary constraints $\Pi=0$ in \bref{ADDs1153a.4S} become, in terms of the new momenta, as
\be
\pi_\phi=\pi_{\phi'}=\pi_{\vp}=\pi_{\7\vp}=0.
\label{ADDnewpiconst}\ee
The gauge $SO(p+1)$ and $SO(D-(p+1))$ degrees of freedom can be eliminated 
by imposing gauge fixing constraints for the $ SO(p+1)$ and  
$ SO(D-(p+1))$ matrices. The simplest choice for this gauge fixing is 
\be
{\phi_a}^b={\D_a}^b,\qquad {\phi_{a'}}^{b'}={\D_{a'}}^{b'}.
\label{GF}
\ee 
Note now that $\pi_{\7\vp}=0$ and $\7\vp-\vp=0$ are the second class constraints 
to eliminate the $(p+1)(D-(p+1))$ canonical pairs 
\be
({\7\vp_a}{}^{b'}, {\pi_{\7\vp}^a}_{b'})=({\vp_a}^{b'}, 0).
\ee
Using them $\Phi$ is expressed in terms of $\vp$ only as
in \cite{Gomis:2006xw}, see also \cite{Kugo:1999mf},
 \bea
{\Phi_\ua}^\ub&=&
\begin{pmatrix}
{B_{1a}}^{b} & {{\vp}_a}^{~c'}{B_{2c'}}^{b'} \cr 
-{\vp^c}_{a'}{B_{1c}}^{b}  & {B_{2a'}}^{b'} \end{pmatrix}. 
\label{ADDPhiparagf}\eea

The $\vp$'s are further determined by solving the second class constraints ${\chi_m}^{a'}=0$ and the  
$(D-(p+1))$ constraints ${\7\phi}^{a'}=0$. 
In the gauge \bref{GF} they are
\be
{\chi_m}^{a'}=(\pa_mx^a{\vp_a}^{b'}+\pa_mx^{b'}){{B_2}_{b'}}^{a'}=0,
\ee 
\be
{\7\phi}^{a'}=(p^a{\vp_a}^{b'}+p^{b'}){{B_2}_{b'}}^{a'}=0.
\ee 
They are combined with $\pi_\vp=0$  in \bref{ADDnewpiconst} to form $(p+1)(D-(p+1))$ second class constraints pairs
to eliminate ${\vp_a}{}^{b'}$ in terms of $p_\ua$ and $\pa_mx^\ua$ as in \bref{sol}.
Notice that 
 ${\vp_a}{}^{b'}$ are written as functions of the 
phase space variables of the $p$-brane. 

Now all second class constraints are used to reduce the phase space to that of $(x^\ua, p_\ua)$. 
There remain the first class constraints   $\CH_\perp=\CH_m=0$, 
and the first class Hamiltonian becomes 
\be
\CH=\int d^p\xi\,(\hat\lam^\perp\,\CH_\perp^1+\hat\lam^m\,\CH_m^1),
\label{ADDDhamiltonian4}\ee
 \be
\CH_\perp^1=\frac1{2\kappa}(\h^{\ua\ub}p_\ua p_\ub+\kappa^2\,{\bf g}),\qquad
\CH_m^1=p_\ua\,\pa_m x^\ua. \label{ADDDiffgenerator20}\ee
The action of $p$-brane obtained from the non-linear realization in the reduced space leads
to the canonical action of a Dirac-Nambu-Goto $p$-brane action in configuration space.


\end{document}